\documentclass[12pt,a4paper]{article}
\usepackage[dvips]{graphics}

\newcommand{\ybox}[2]	{
 \begin{center}
 \resizebox{!}{#1\textheight}
{\includegraphics{#2.eps}}
 \end{center}		}

\newcommand{\twobox}[3]	{
 \begin{center}
 \resizebox{!}{#1\textheight}
{\includegraphics{#2.eps} \includegraphics{#3.eps}}
 \end{center}		}

\begin{document}
\thispagestyle{empty}

\begin{center}

{\LARGE \bf Constraints on the Ultra High Energy Photon flux using 
inclined showers from the Haverah Park array}  
\end{center}

%
\begin{center}
{\bf  M. Ave$^1$, J.A.~Hinton$^{1,2}$, R.A.~V\'azquez$^3$, \\
A.A.~Watson$^1$, and E.~Zas$^3$}\\
$^1$ {\it Department of Physics and Astronomy\\
 University of Leeds, Leeds LS2 9JT ,UK \\}
$^2$ {\it Enrico Fermi Institute, University of Chicago, \\ 
5640 Ellis av., Chicago IL 60637, U.S.\\}
$^3$ {\it Departamento de F\'\i sica de Part\'\i culas,\\
Universidad de Santiago, 15706 Santiago de Compostela, Spain\\}
\end{center}

\begin{abstract}
We describe a method to analyse inclined air showers produced by ultra high energy 
cosmic rays using an analytical description of the muon densities. We report 
the results obtained using data from inclined events 
($60^{\circ}<\theta<80^{\circ}$) recorded by the Haverah Park shower detector 
for energies above $10^{19}~$eV. Using mass independent knowledge of the UHECR 
spectrum obtained from vertical air shower measurements and comparing the 
expected horizontal shower rate to the reported measurements 
we show that above $10^{19}$ eV less than $48\%$ 
of the primary cosmic rays can be photons at the $95 \%$ confidence 
level and above $4 \times 10^{19}$ eV less than $50\%$ of the cosmic rays can be 
photonic at the same confidence level. These limits place important constraints 
on some models of the origin of ultra high-energy cosmic rays. 
\end{abstract}

\section{Introduction}
\label{intro.sec}

The question of the origin of cosmic rays (CRs) of the highest energies is
currently a subject of much intense debate and discussion. The highest energy
cosmic ray ($3 \times 10^{20}$ eV) was detected by Fly's Eye fluorescence
detector \cite{FlysEye} confirming the existence of cosmic rays with
macroscopic energies above the Greisen, Zatsepin, and Kuz'min cut-off ($4
\times 10^{19}$~eV) \cite{GZK}. In addition the AGASA group have reported 8
events with energies above 100 EeV and other very energetic events with
energies beyond the GZK cut-off have been described by the Volcano Ranch,
Haverah Park and Yakutsk groups \cite{Vranch,yakutsk,HPevents}. These ultra
high-energy cosmic rays (UHECR) pose a serious challenge for conventional
theories of CRs based on stochastic acceleration. The non-observation of the
high energy cut-off expected because of the interactions with the Cosmic
Microwave Background (CMB), indicates that their sources must be nearby thus
posing serious restrictions as to their origin. There is currently a
significant experimental effort underway, focussed around HiRes \cite{HiRes}, 
the Pierre Auger Observatory \cite{auger} and EUSO \cite{EUSO}, 
aimed at dramatically improving the statistics at the highest energies. 

The old idea of attempting to detect high-energy neutrinos through studying
very inclined air showers (HAS)\cite{Berezinskii} has been recently revived
with the calculation of the acceptance of the Auger Observatories for the
detection of high-energy neutrinos \cite{Capelle}. Ultra High Energy (UHE)
neutrinos (above EeV) are almost inevitable in models that seek to
explain the UHE cosmic rays. At large zenith angles, cosmic rays (whether they
are protons, nuclei or photons) develop ordinary showers in the top layers of
the atmosphere in a very similar fashion to the well-understood vertical
showers. Their electromagnetic component is, however, almost completely
absorbed by the greatly enhanced atmospheric slant depth (3000 g~cm$^{-2}$ at
70$^\circ$ from zenith) and thus prevented from reaching ground level. 
High energy neutrinos may induce HAS much deeper in the atmosphere 
close to an air shower array. By contrast, these showers at ground level 
resemble vertical air showers in their particle content and other features. 

The main background to UHE neutrino induced HAS is expected to be due to the 
remaining muon component of the cosmic rays showers, after practically all of 
the electromagnetic component is absorbed. These $muonic~showers$ that 
penetrate the whole atmospheric depth to ground level are the object of this 
study. Although originally this project was conceived as a study of the 
background to neutrino-induced showers we have come to the conclusion that the 
interest in HAS induced by cosmic rays goes well beyond that expectation. The 
measurement of high zenith angle showers will enhance the aperture of the 
existing air shower arrays, and will increase the data on cosmic ray arrival 
directions to previously inaccessible directions in galactic coordinates 
\cite{utah99}. Besides these obvious advantages, high zenith angle cosmic ray 
showers are unique because the shower front is dominated by relatively energetic 
muons that travel long distances, opening up the possibility 
of probing interactions in a region of phase space quite inaccessible in 
vertical air showers.

Cosmic ray induced HAS are different from vertical showers mainly because they 
consist largely of muons which are produced far from ground level.  The 
particle density profiles for HAS induced by protons or heavy nuclei display 
complex muon patterns at the ground which result from the long path lengths 
traveled by the muons in the presence of the Earth's magnetic field 
\cite{Hillas,Andrews}.  These patterns are difficult to analyse 
\cite{Antonov98} and invalidate the conventional approach used for 
interpretation of low zenith angle showers ($< 60^\circ$), which is usually 
based on the approximate circular symmetry of the density profiles. The analysis 
of HAS produced by cosmic rays requires a radically different approach
such as the one presented in this work.

We apply our approach to data recorded by the Haverah Park experiment. The 
Haverah Park array, being made of 1.2~m deep water \v Cerenkov tanks 
\cite{haverah}, is the detector array so far constructed which is best suited on 
geometrical considerations for the analysis of very large inclined showers. 
Moreover it can be considered as a prototype of the Auger Observatories, which 
will employ water \v Cerenkov tanks of identical depth. The quantitative 
aspects of our results are very specific to the water \v Cerenkov technique as 
we have previously taken into account in great detail the interaction of the 
shower particles in the water detectors \cite{rate}.

In this paper we give a much more detailed account of a report 
already published \cite{PRL}. 
The present work is organized as follows: In section 2 we discuss the main
features of inclined showers, the muon distributions, the different sources of
electrons and photons and the shower front curvature. 
In section 3 we give a brief description of the Haverah Park array, its
detectors and their response to the passage of different particles from 
the shower front. 
In section 4 we develop an algorithm reconstruct the arrival
directions and energies of inclined air showers detected with Haverah Park.
In section 5 we describe a procedure to generate artificial events based on
shower generation and measurements of the cosmic ray spectrum.  In section 6
we compare the high zenith angle data to the artificial event distributions
obtained under different assumptions about the nature of the primary particles
that constitute the cosmic ray energy spectrum above $10^{19}$~eV. We 
extract bounds on photons above $10^{19}$ eV. Finally 
in section 7 we discuss our results and review their implications. 
In a subsequent paper we will describe the use of this technique
to yield the proton/iron ratio as a function of energy above 1 EeV. 

\section{Inclined air showers}
\label{sim.sec}

As the zenith angle varies from the vertical, $\theta=0^\circ$, to the
horizontal, $\theta=90^\circ$, direction, the slant matter depth rises from
$\sim 1000$ to $\sim 36000$~g~cm$^{-2}$ and for angles above $60^\circ$ the
cosmic ray showers at ground level are observed well past the shower maximum.
In inclined showers of zenith angles exceeding $70^\circ$ the 
electromagnetic component arising from the hadron shower through $\pi^0$ 
decay can be neglected at ground level. For zenith angles 
between $60^\circ$ and $70^\circ$ the relative signal of this component 
in a 1.2~m deep water \v Cerenkov detector is small except for 
distances within a few hundred meters from shower axis. 
While the electromagnetic component of air showers
is exponentially attenuated with depth, the muons that are too energetic to
decay, have few catastrophic interactions and only suffer ionization losses,
scattering and geomagnetic deflection. They constitute the dominant component
of the shower front for inclined showers. The muon patterns at ground level 
have been studied in \cite{model}.  
There is a residual electromagnetic component in the
shower front which is produced by the muons themselves, mostly through muon
decay. Other muon interactions contribute either little to the electromagnetic
component or only within a narrow region about shower axis.

\subsection{The muon component}
\label{muonshowers}

The distribution of the muon component at ground level 
becomes complex at large zenith angles because of magnetic field effects. 
The spatial distribution of muons can no longer be characterized by a simple 
function of one parameter (distance to the shower axis, $r$) because of the 
asymmetry generated mostly by the geomagnetic effects. In 
\cite{model} an analytic model to account for the average muon number 
densities 
at ground level in presence of a magnetic field for proton showers at high 
zenith angles is presented and described in detail. We outline its main 
features because the work presented here relies heavily upon it. 

The approach consists of studying the muon distributions in the absence of 
magnetic field effects so that they have cylindrical symmetry to an excellent 
approximation. The distributions are described by functions of one 
variable ($r$), the lateral distribution functions, in a plane perpendicular 
to the shower axis, the transverse plane. 
A very strong anticorrelation between the average muon energy and 
distance of the muon from the shower axis has been described \cite{model}. 

Magnetic deviations of the muons are subsequently applied to the 
circularly symmetric distributions, making use of the aforementioned 
anticorrelation and assuming the muons are produced in a fixed region of 
the atmosphere. 
The magnetic distortions induced in the muon distributions are 
described by considering the projection of the Earth's magnetic field onto 
the transverse plane. As the zenith angle increases the patterns obtained 
in the transverse plane gradually change from elliptical distributions to 
two lobed figures reflecting an increased distance traveled by the muons 
which results in enhanced distortions. The double lobe patterns correspond 
to positive and negative muons totally separated by the magnetic field 
which acts as a spectrometer for the muons in the shower. Moreover 
as the azimuth changes, both the magnitude of the magnetic field projection 
onto the transverse plane and its relative orientation 
with respect to the ground, change, further increasing the diversity of 
the resulting patterns projected onto the ground.   

The description of the average muon density patterns thus 
requires three inputs:

$\bullet$ The lateral distribution function (LDF).

$\bullet$ The average muon energy as a function of radius (${\cal E}(r)$).

$\bullet$ The mean distance to the muon production point ($l_0(\theta)$).

\noindent All these values must first be evaluated in the absence of magnetic 
effects.  The model also requires knowledge of the muon energy distribution at a fixed distance to the shower axis. A log-normal distribution of width 0.4 
has been found to be sufficiently accurate for all practical purposes.

\begin{figure}
\ybox{0.4}{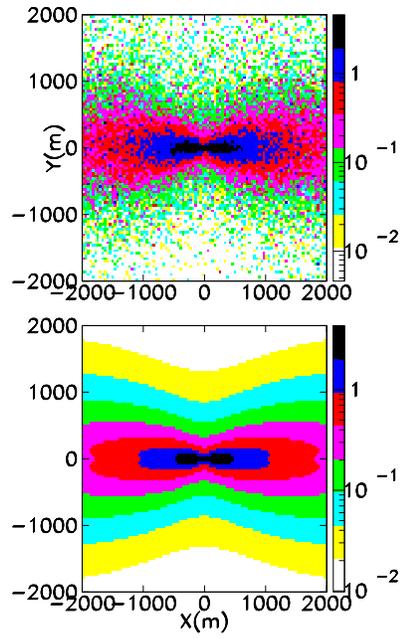}   
\caption{Contour plots of the muon density patterns in the transverse plane 
for $10^{19}$ eV proton showers with an incident zenith angle of $\theta=80^\circ$ 
and azimuth angle $\phi=0^\circ$ as obtained in the simulation 
(upper panel) and with the analytic approach described (lower panel).}
\label{contour80}
\end{figure}

The validity of the analytical description has been evaluated by comparing
full shower simulations for different arrival directions with those obtained
by this procedure.  A comparison of muon densities in this model to those
obtained by simulation is shown in Fig. \ref{contour80}.  The simulations of
the density distributions both with and without the magnetic field have been
made with the AIRES \cite{AIRES} code. Tests for proton showers using the 
SIBYLL \cite{sibyll} hadronic interaction
generator at a fixed energy of $10^{19}$~eV, and four different zeniths are
described in \cite{model}.
 
This approach is independent of model details and can be applied to other
hadronic generators, mass compositions, and energies by changing the
corresponding inputs.  It allows the comparison of muon density patterns at
ground level through these simple inputs.  A significant advantage is that,
provided the lateral distribution function is parameterized by a continuous
function, the muon density patterns obtained in the transverse plane are
smooth functions in contrast to distributions obtained with any Monte Carlo
simulation.  This key point allows us to reconstruct the energy of individual
events, as we describe below.

\begin{figure}
\ybox{0.35}{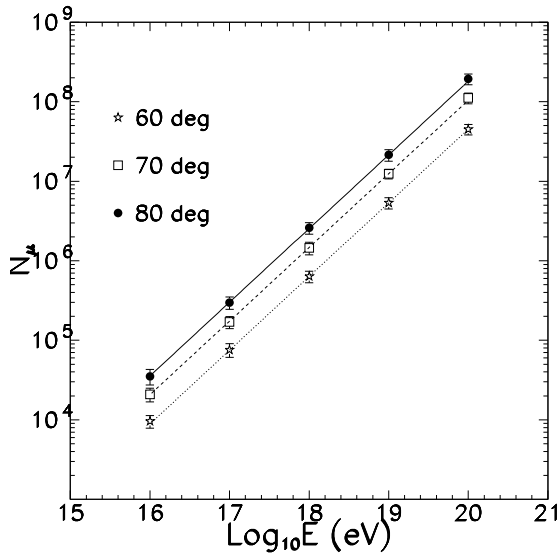}   
\caption{The relationship of total muon number to primary energy 
for protons of four zenith angles using the QGSJET model \cite{QGSJET}.}
\label{escale}
\end{figure}

We have also found \cite{rate} that to a good approximation the inputs to our 
model are energy independent for a given primary, so the muon number density
distributions for any primary energy can be obtained
simply by normalizing the total number of muons of a fixed energy shower. 
These results apply to showers both with and without the magnetic field. The 
energy dependence of the normalization factor can be obtained by monitoring the 
total muon number in the showers ($N_{\mu}$). The values of $N_{\mu}$ from 
simulations 
are plotted in Fig.~\ref{escale} 
for four different zenith angles. The energy dependence of the normalization 
can be taken into account accurately by a simple relation of the following form:
\begin{equation}
N_\mu=N_0~\left[{E \over 10^{19}~{\rm eV}}\right]^{\beta}, 
\label{Escaling}
\end{equation}
where $N_0$ (the number of muons at $10^{19}$~eV) and $\beta$ are constant 
parameters for a given hadronic interaction model and mass composition. For 
different zenith angles the energy scaling index, $\beta$, is the same and only 
the normalization $N_0$ changes. 

\begin{figure}
\ybox{0.3}{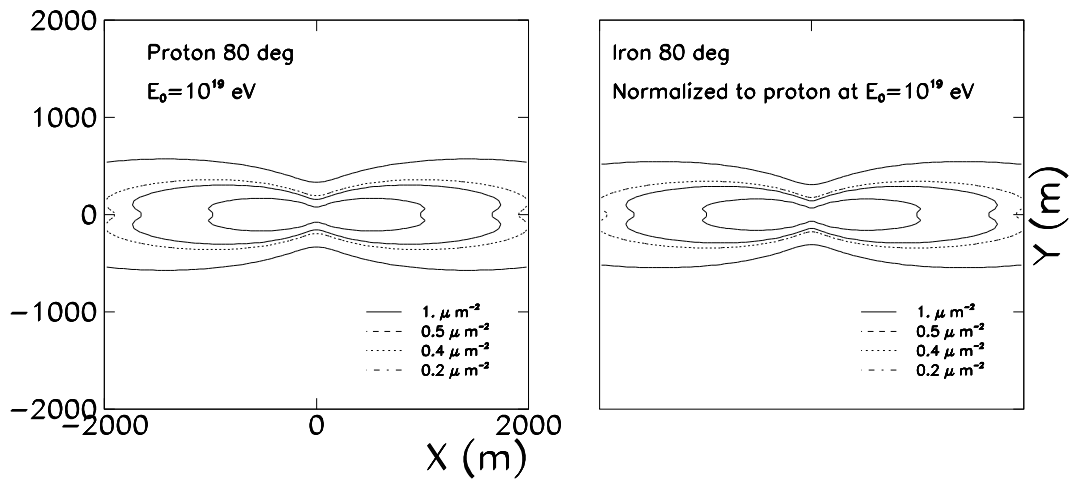}  
\caption{Muon density patterns in the transverse plane for a $10^{19}~$eV 
proton shower incident with $80^\circ$ zenith angle, as well as for an 
iron shower normalized to the same number of muons for comparison.}
\label{fevsp2}
\end{figure}

\begin{figure}
\ybox{0.5}{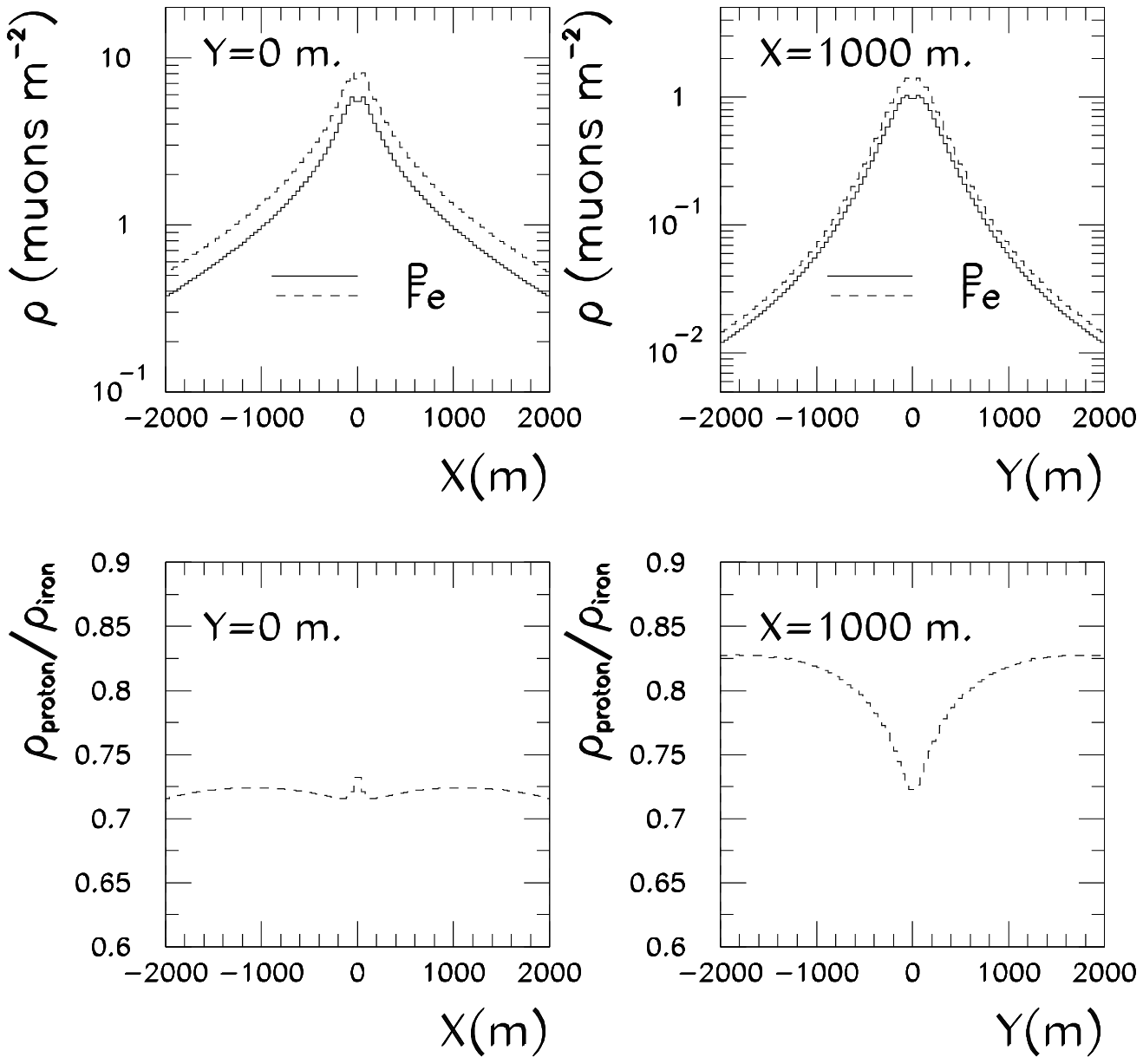} 
\caption{Top left panel: Muon number density in the transverse plane as a function
of the $x$ coordinate for a fixed value of $y$ as obtained with the model
for a proton (full line) and iron (dashed line) shower of $10^{19}~$eV arriving 
with 80$^\circ$ zenith angle. The $y$-axis is chosen parallel to the 
magnetic North. Densities are calculated in 40~m x 40~m bins and the $x$
axis bins shown are centered at $y=0$. 
Bottom left panel: ratio of for proton and iron densities along the $x$ axis. 
Top right panel: Muon number density taking $y$ bins for a fixed value of 
$x$= 1000 m. Bottom right panel: density  ratio along the $y$ coordinate.}
\label{fevsp}
\end{figure}

Furthermore the muon distributions at ground level are hardly different in
shape for iron and proton. This is illustrated in Figs.~\ref{fevsp2} and \ref{fevsp}
where muon densities patterns and densities along given lines parallel to the
$x$ and $y$ directions in the transverse plane axes are compared for iron and
proton primaries. To a good approximation the differences can be accounted for by
differences in the total number of muons.  For a given model and primary
composition the energy dependence of very inclined showers can be
parameterized with only two parameters. In Table~\ref{nmu.tab} the results for
these two parameters for proton and iron in two interaction models are shown.
\begin{table}
\begin{center}
\begin{tabular}{|lrcc|} \hline
Model  & A  & $\beta$ & $N_{\mu}$ ($10^{19}$ eV) \\\hline\hline
SIBYLL & 1  & 0.880 & 1.6~10$^{7}$ \\ 
       & 56 & 0.873 & 2.2~10$^{7}$ \\  \hline
QGSJET & 1  & 0.926 & 2.1~10$^{7}$ \\  
       & 56 & 0.909 & 2.8~10$^{7}$ \\  \hline
\end{tabular}
\end{center}
\caption{Relationship between muon number and primary energy for 
different models and primary masses (see Eq. (\ref{Escaling})), for
 a zenith angle of 60$^\circ$.}
\label{nmu.tab}
\end{table}

\begin{figure}  
\ybox{0.7}{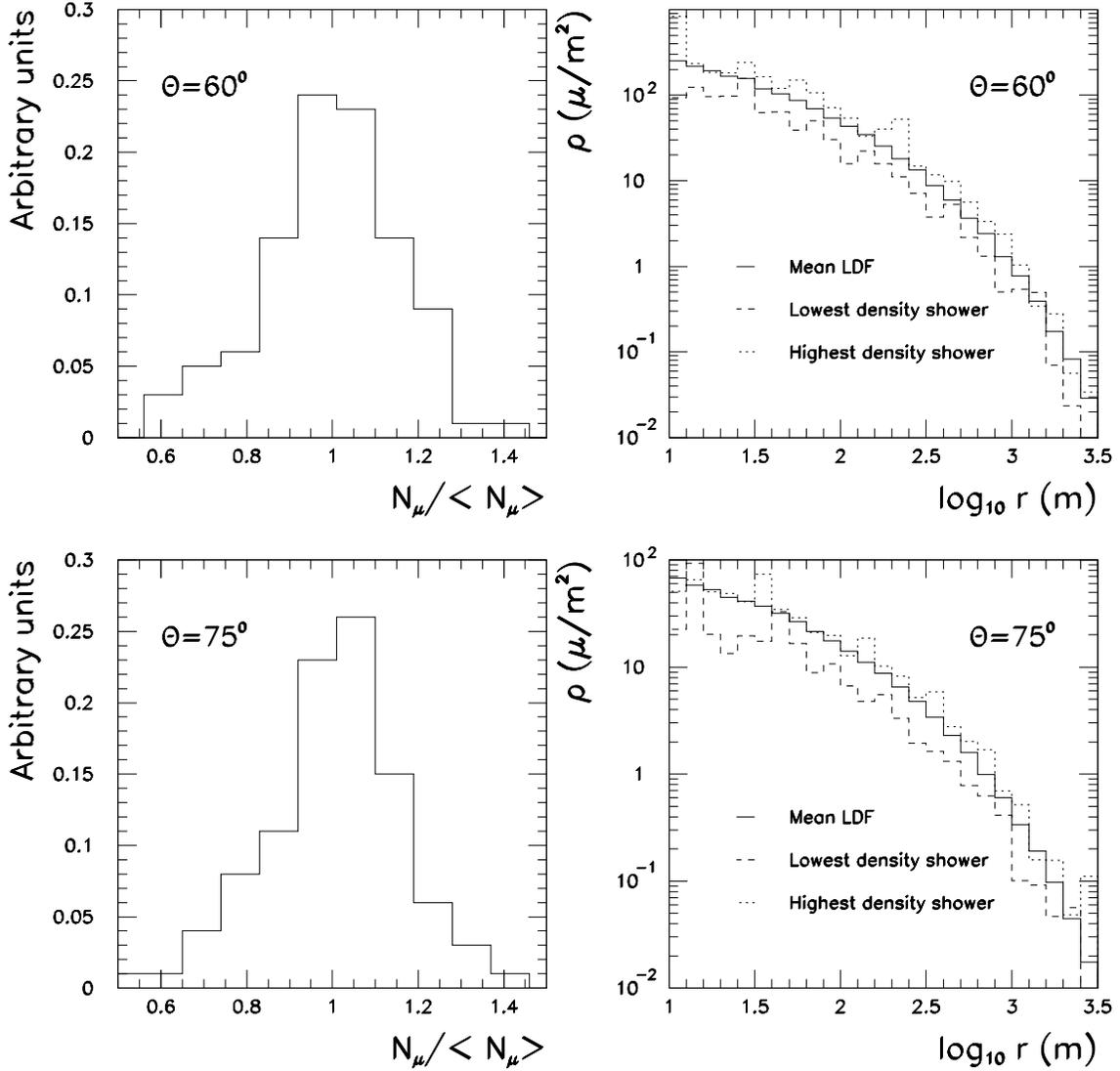}  
\caption{Left panels: Distribution of the total number of muons for 100 individual 
proton showers of energy $10^{19}~$ eV and zenith angle $60^\circ$ (top graphs), 
$75^\circ$ (bottom graphs), simulated with AIRES code and QGSJET hadronic generator. 
Right panels: Muon lateral distribution for the extreme showers with largest and 
lowest number of muons compared to the mean.}
\label{fluct}
\end{figure}

It is well known that fluctuations in shower development can enhance the
trigger rate for air showers produced by lower energy primaries because of the
steep cosmic ray spectrum. The fluctuations to larger numbers of particles
allow some of the more numerous low energy showers to trigger the detector. We
have also studied muon number fluctuations at ground level and how they relate to
shower development (mean muon production height) and average muon energy. We
have found that the mean muon energy correlates strongly with production
height but that most of the number density fluctuations can be accounted for
by fluctuations in muon number. Fluctuations in the total number of
muons are mainly due to fluctuations in the depth of maximum, which are related 
to fluctuations in the first interaction depth, as well as to fluctuations in 
the neutral to charged pion ratios in the first interactions. In
Fig. \ref{fluct} the distribution of the muon number for a set of 100
showers with the same primary energy and mass composition is plotted.
Although the distribution is slightly asymmetric with a tail towards low
$N_\mu$ number, in this work we have assumed a gaussian distribution with a
width of $\sigma_{N_\mu}=0.2 <N_{\mu}>$.  In Fig. \ref{fluct} we compare 
the mean muon
density as a function of $r$ to that of the extreme cases of muon rich and
muon poor showers obtained in the simulation. No significant changes in the
shape of the muon LDF need to be considered for distances beyond about 100 m.

\begin{figure}
\ybox{0.3}{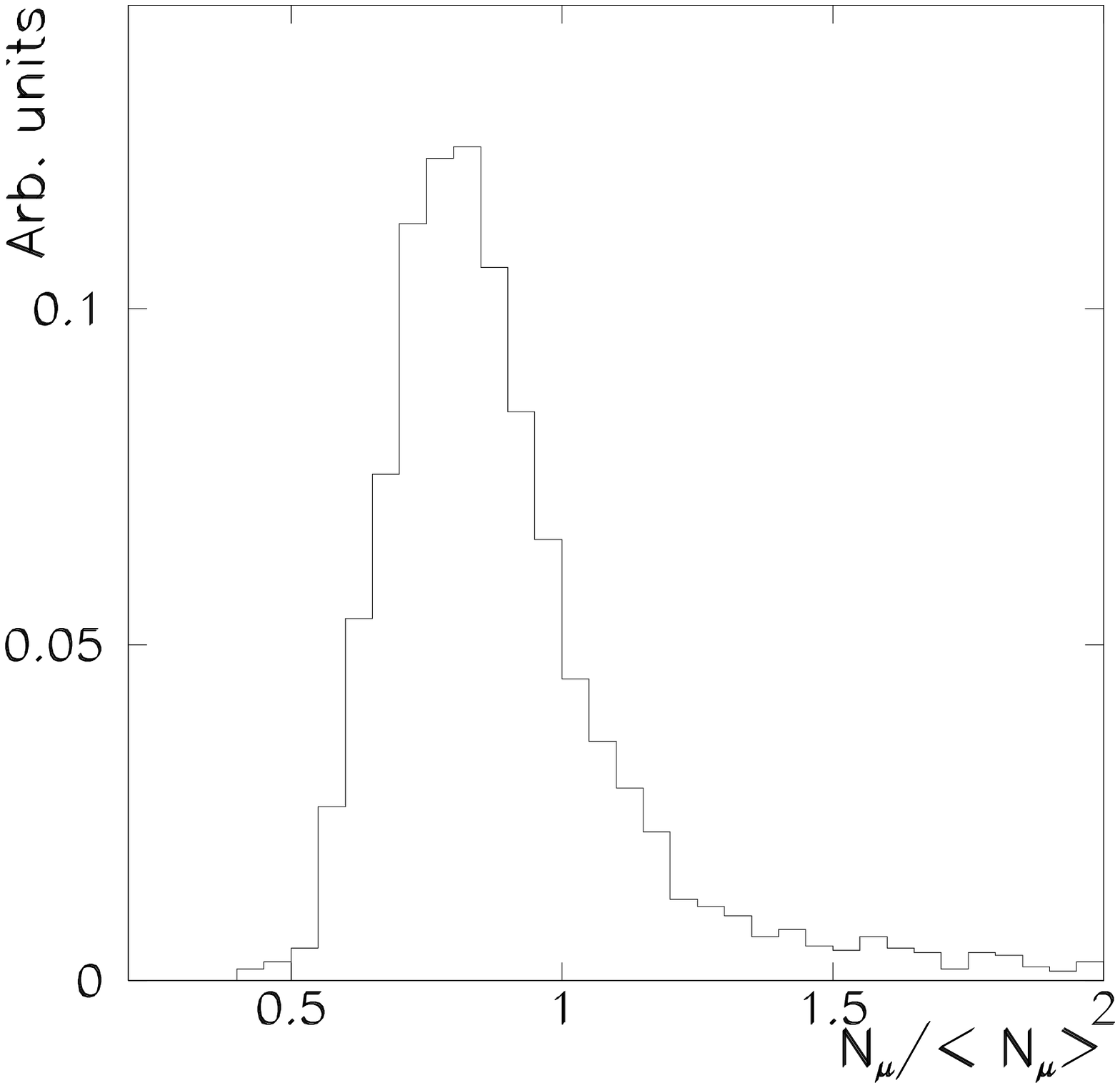}  
\caption{Distribution of number of muons for individual photon showers at
$10^{19}~$ eV simulated with AIRES code and QGSJET hadronic generator. This
plot was obtained combining different zenith angles normalizing the number of
muons of each individual shower to the mean value at a given zenith angle.}
\label{fluct_ph}
\end{figure}

Fluctuations in the number of muons in photon-induced showers are rather 
different from those in hadron-induced showers. If the first interaction of the 
incident photon happens to be hadronic (probability R$\sim$ 0.01 at $10^{19}$ 
eV) then the shower is indistinguishable from a hadronic shower. For the 
distribution of the total number of muons in a photon shower we can therefore 
expect a long tail of showers with large number of muons, as can be seen in 
Fig.~\ref{fluct_ph}.  

\subsection{The electromagnetic component of very inclined showers}

As will be described in the next section a detector that uses water 
\v Cerenkov tanks is more efficient for detecting muons than electrons and photons 
because muons typically go through the whole tank and thus give larger signals in 
the tanks than the typically lower energy electrons and photons. 
The electromagnetic component of inclined shower induced by a proton or a nucleus 
has been studied with the help of both analytical calculations and Monte Carlo 
simulations using the AIRES code \cite{AIRES}. 
We can distinguish three components according to their origin:   
\begin{figure}
\ybox{0.7}{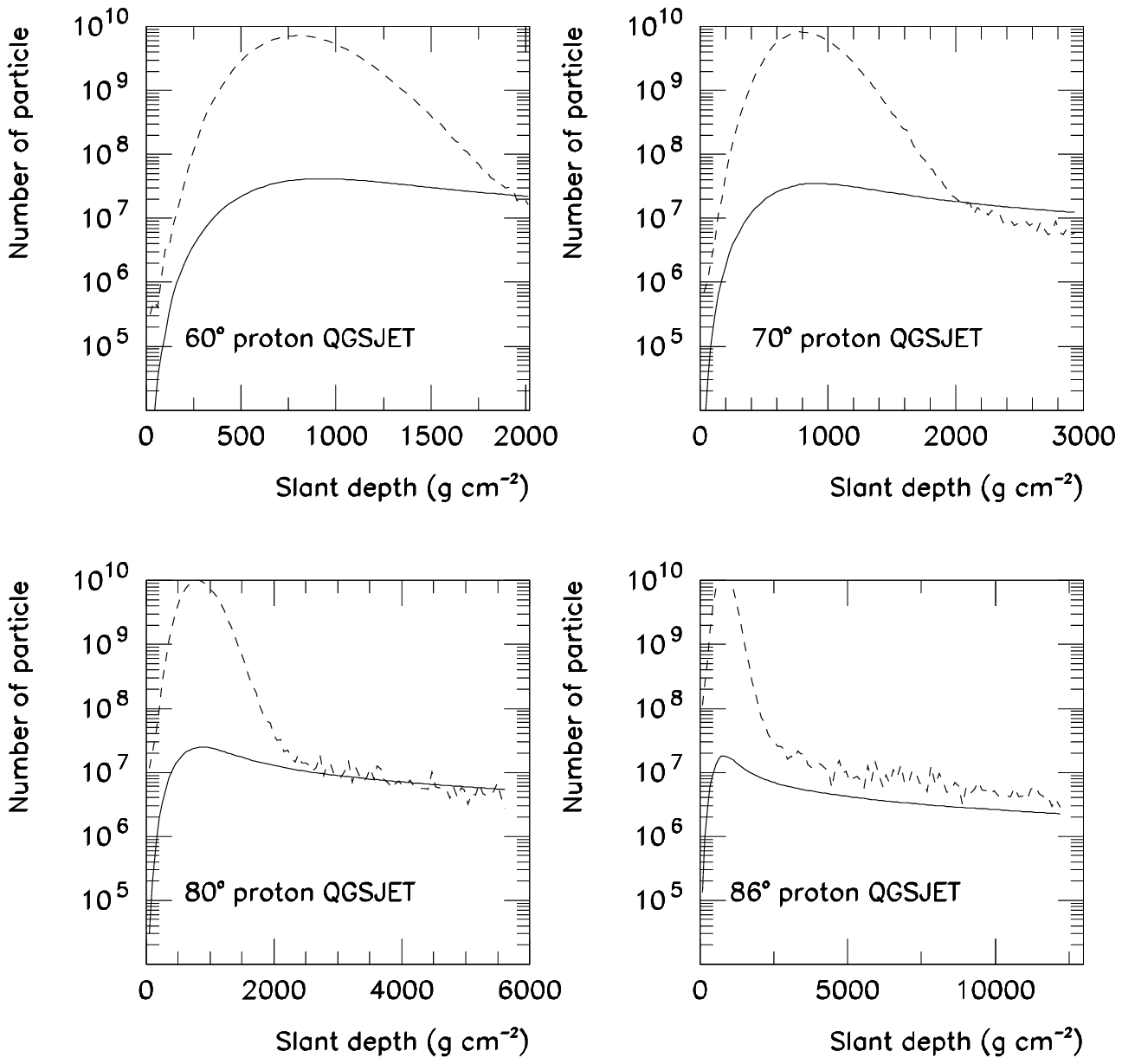}  
\caption{The average longitudinal development of the muon (continuous line) 
and electron (dashed lines) components for 100 proton showers of energy 
$10^{19}$~eV and different zenith angles. At depths exceeding 3000 
g cm$^{-2}$, or equivalently for zenith angles greater than 
70$^{\circ}$, the electromagnetic component is mainly due to 
muon decay.}
\label{shower}
\end{figure}

$\bullet$ {\it The component fed by muon decay}: The longitudinal developments
of the electron and muon components are shown in Fig.~\ref{shower} for
$10^{19}~$eV proton showers arriving with four different zenith angles.  In
these simulations the effects of muon bremsstrahlung, pair production and
nuclear interactions are not included.  The most striking feature of these
figures is that after reaching shower maximum there is a residual component
that follows closely the muon depth distribution. This effect is mostly due to
electrons from muon decay. The relative number of electromagnetic particles
(electrons and photons) with respect to the muons is seen to be practically
independent of depth and only mildly increasing with zenith angle.

\begin{figure}
\twobox{0.28}{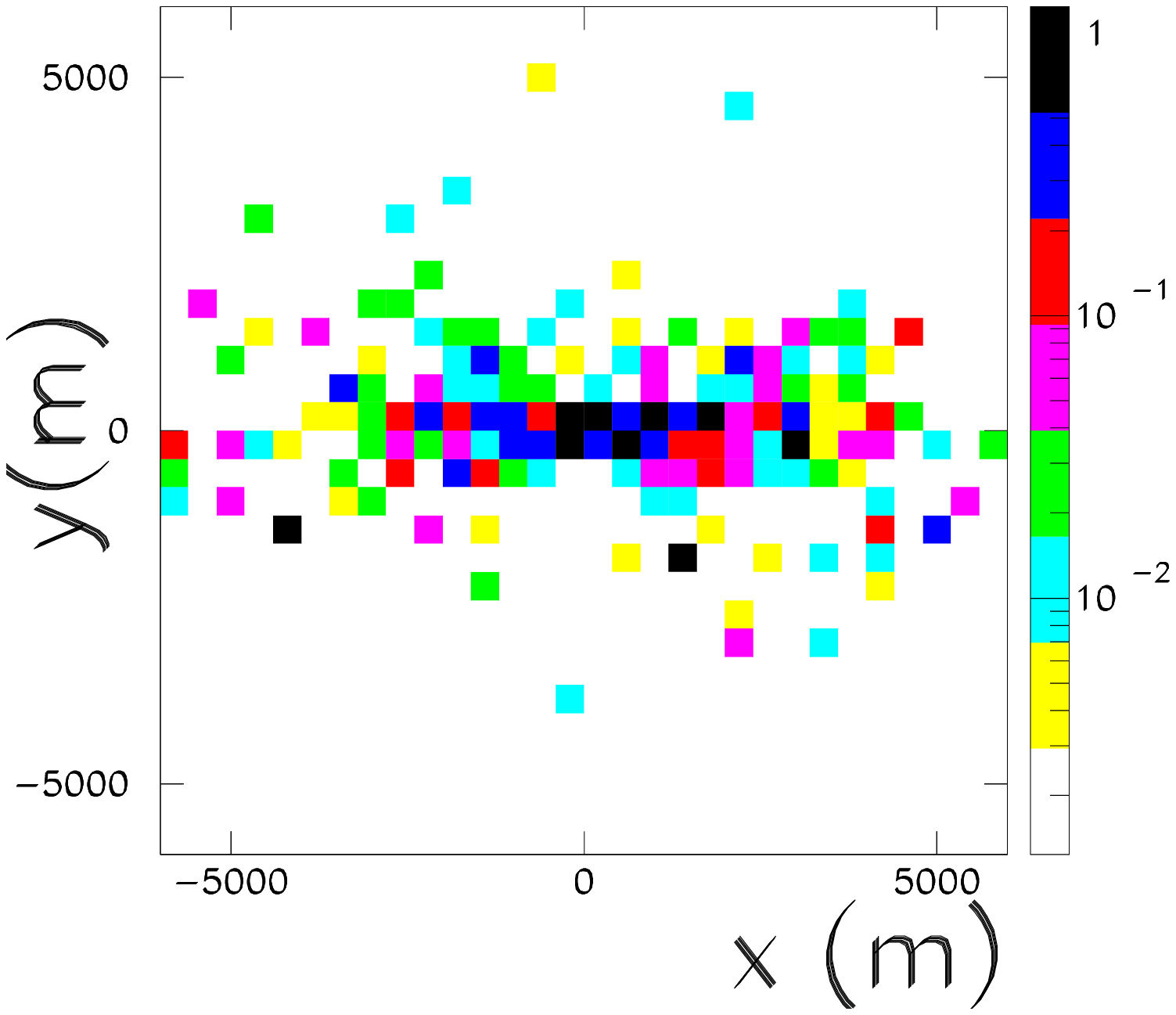}{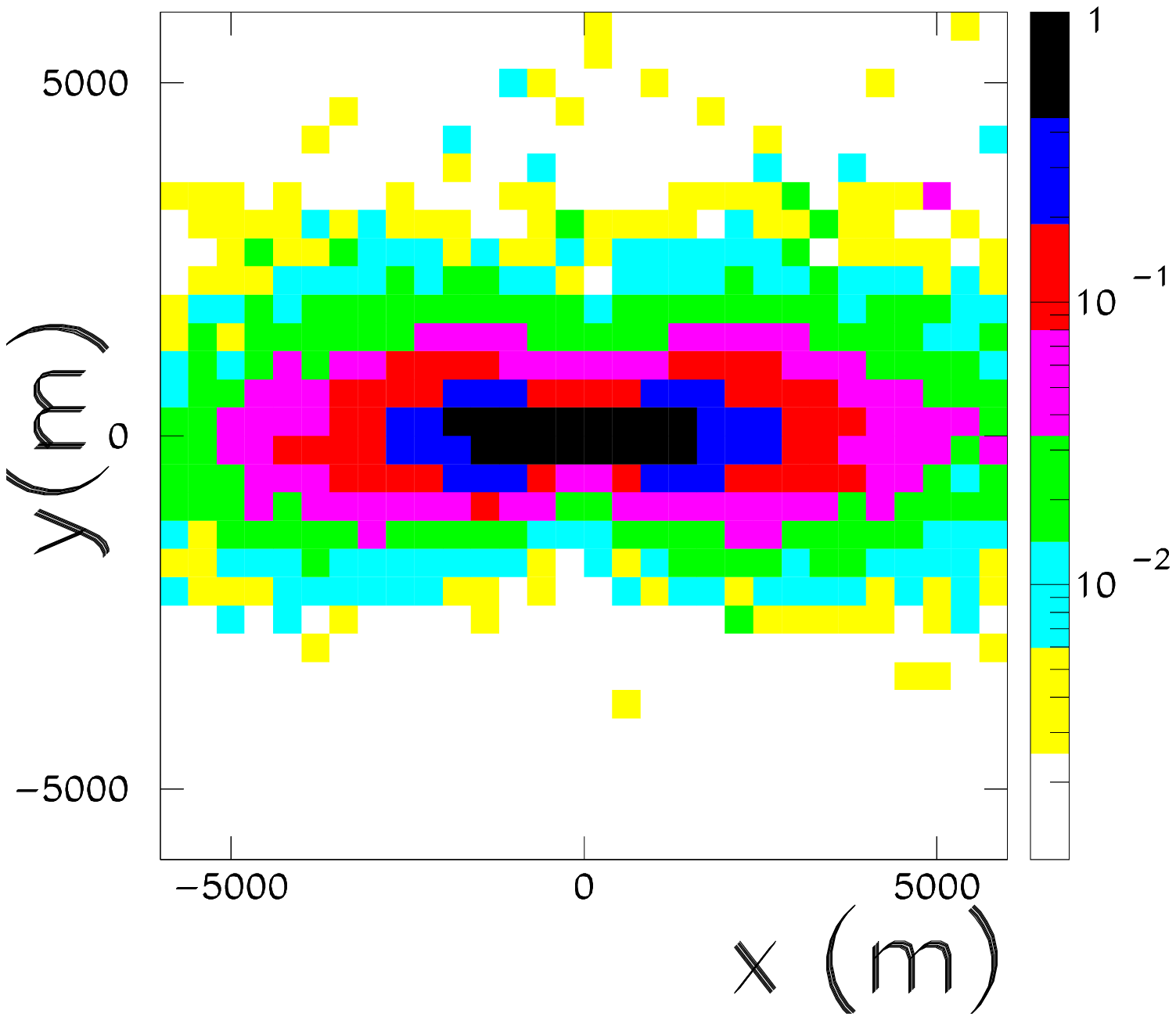}  
\caption{Left: Density pattern of the muons in the transverse plane for
a $10^{19}$~eV proton induced shower with a zenith of $80^\circ$, and including 
geomagnetic effects. Right: Electron density pattern for the same shower.}
\label{el_prof}
\end{figure}

As electrons and photons develop from multiple electromagnetic subshowers 
their energy distribution is essentially the same as that of a typical  
air shower.
The ratio fluctuates because of the discreteness of the energy deposition. 
The lateral distribution follows that of the muons rather closely, as shown in
Fig. \ref{el_prof} unlike the LDF for electrons in near vertical air showers.

$\bullet$ {\it The component fed by $\pi^0$ decay}: 
Fig.~\ref{shower} clearly shows an early electromagnetic part mostly induced by the
$\pi^0$'s from the hadronic interactions which decay into photons that
cascade down the atmosphere. This component becomes exponentially suppressed
after shower maximum and is quite unimportant for inclined showers. 
Indeed even at $60^\circ$ the electromagnetic component of a
$10^{19}$~eV proton shower which can be directly associated to $\pi^0$
decay is already low and confined within a relatively small region of
about 200~m around shower axis. 
For $\theta>70^\circ$ we do not have a significant contribution 
to the electromagnetic component from $\pi^0$ decay. 

$\bullet$ {\it The component fed by muon interactions (bremsstrahlung, pair 
production and muon nuclear interactions)}: The muons in very inclined air 
showers have greater energies and traverse more matter than in the vertical case 
so these processes need to be considered. We have estimated the global 
bremsstrahlung contribution by considering the muon energy spectrum of a single 
shower, folding it analytically with the bremsstrahlung cross-section and the 
Greisen parameterization, see \cite{ZHV}. For an $80^{\circ}$ zenith and 
$10^{19}~$eV proton shower the total number of electrons and positrons ($N_e$) 
obtained is about $2.5~10^{4}$. These are mostly due to the muons in the energy 
range between 30~GeV and 500~GeV. This component arises also from 
electromagnetic sub-showers and its energy distribution should also reflect that 
of electromagnetic cascades.  If we multiply, conservatively, the total number 
of electrons by a factor three to account for the two other muon interactions, 
it is still a factor of $\sim$50 below the total number of muons in the shower, 
and negligible compared to the electromagnetic contribution from muon decay.
This component has recently been incorporated into a new version of AIRES and 
analysed fully with simulation in \cite{cillis}. These results show that the 
electromagnetic component dominates over the muons only for distance to shower 
axis below $\sim 100$~m in agreement with our calculations. 
%

To evaluate the relative importance of each of these contributions to the shower 
front relative to the signal induced by the muons in inclined 
showers, we need to weight the different type of particles at ground with the signal 
produced by them in a given experiment. In  section 3 it will be shown that 
the quantitative effects of the electromagnetic component for water \v Cerenkov tanks 
such as those used in the Haverah Park array are unimportant except for distances very 
close to the shower axis. 

\subsection{ Shower front curvature }
\label{tonto}

Very inclined air showers detected at ground level are mostly dominated by muons 
which travel long distances without large attenuations, as discussed in the 
previous section. We expect the curvature and the time spread of the muon front 
to be smaller than in vertical showers. 
We have studied the arrival time of the muons through simulations performed with 
AIRES code. We have simulated 100 proton induced showers at $10^{19}$~eV for 
three different zeniths ($60^\circ$, $70^\circ$, and $80^\circ$).  
The output from the simulation gives the arrival time of the muons at ground 
level, but we prefer to study the shower front (thickness and shape) in the 
transverse plane. We have projected the muons onto the transverse plane, 
correcting the arrival times at the ground with the different muon paths to 
reach this plane. 
After this correction we get the time distributions of the 
muons for different bins in distance to the shower axis. 

\begin{figure}
\ybox{0.3}{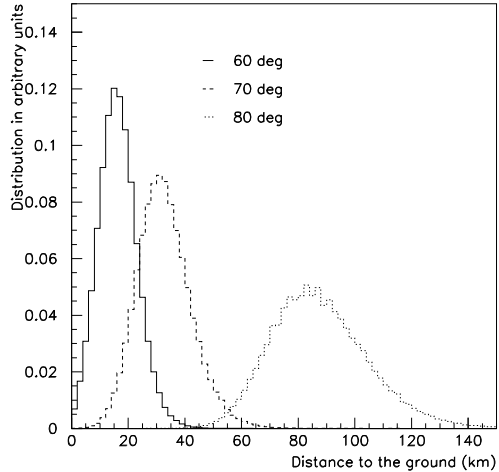}  
\caption{Distribution of distances traveled by the muons from their production
site to the ground for three different zeniths. The ratio of $\sigma/d$ for
the three histograms are 0.4, 0.27 and 0.20 respectively. At 87$^\circ$ (not
shown) the ratio is 0.13.}
\label{hdist}
\end{figure}

The distance traveled by the muons characterizes the most important properties 
of the shower front in inclined showers and is the basis of the analytical 
model discussed in section 2.1. It also characterizes the curvature of the shower 
front. 
Assuming the muons are produced at a fixed point one would expect a 
spherical shower front which turns out to be a fairly good approximation. 
The distributions of distances traveled by the muons from the production site to 
the ground are plotted in Fig. \ref{hdist} for three different zeniths, as 
obtained from simulations. 
The distributions are relatively narrow compared to the mean 
value $<$$d$$>$, so that for a given zenith angle it is a reasonable approximation to 
consider all the muons as coming from a fixed point. 
As the production point is not very sensitive to the nature of the primary 
particle the curvature of the shower front can be also expected to be 
relatively independent of composition. 

Typically the times recorded in a ground array experiment, which are
eventually used for the arrival direction fit, are the relative times of the
onset of the signal at the different detectors. One can visualize the muon
arrival time distribution as the delay associated with the different muon paths
from production to a particular position in the shower front. We take from the 
time distribution the arrival time of the first muon. This
implies that there is another factor that can distort an experimental
reconstruction of the shower front related with the statistical sampling. For
a given number of muons $n$ arriving at a particular detector, we are
effectively sampling the corresponding time distribution $n$ times and then
choosing the earliest time. For a large number of muons, this time will tend
to the geometrical delay of the highest energy muon, but for a small number of
muons the earliest muon will be distributed about a mean value with a width
which decreases with $n$.  As a result there is an additional curvature that
is entirely a statistical effect as was pointed out many years ago
\cite{baxter}.

\begin{figure}  
\ybox{0.6}{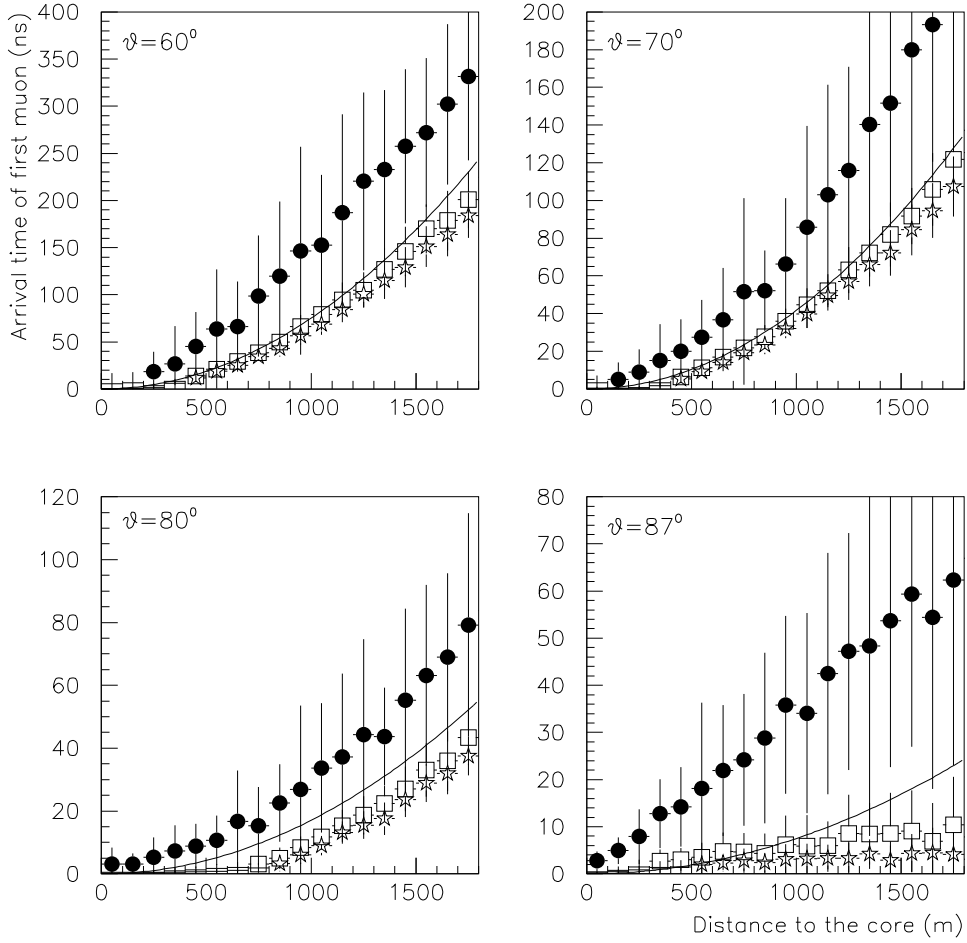}
\caption{Arrival times of the first muon at different distances from the shower axis for a $60^\circ$, $70^\circ$, $80^\circ$, and $87^\circ$ proton induced
 shower, after sampling the time distributions for
 different number of muons.  Dots correspond to a sampling with 1
 muon, open squares to 10 muons, and stars to 100 muons.  The
 continuous line plotted corresponds to a spherical shape with the
 radius of curvature equal to the mean distance traveled by the muons
 to ground ($<d>$).}
\label{curvfront}
\end{figure}

In Fig. \ref{curvfront} we have plotted the arrival time of the first muon in 
the sample for four different zeniths and assuming a different number of muons 
hit the detector.  We have 
superimposed a spherical front with radius of curvature equal to $<d>$. 
The accuracy of this simple approximation seems good enough except for very 
high zeniths. 
As the muon number density drops with the distance from the shower axis, the sampling 
will affect the measured arrival time of the first muons. The curvature  
effectively grows with the distance to the shower axis. This can be accounted for 
as an extra contribution to the error of the measured time. 
In an experimental situation the necessity of spherical 
corrections will be determined by the experimental errors in relation to the 
arrival time delays. We will apply curvature corrections in the event 
reconstruction of inclined showers in section 4. A spherical front assumption 
seems justified except for showers close to the horizontal  when a flat front  
can be assumed because the curvature is very small.

\section{The Haverah Park array}
\label{data.sec}

\begin{figure}
\ybox{0.35}{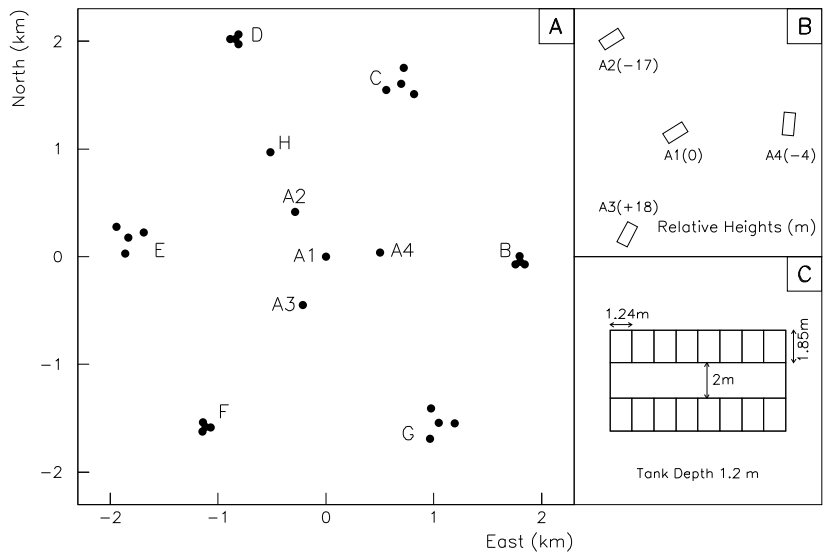}  
\caption{The Haverah Park Array. A) The 2 km array. B) The relative 
heights and orientations of the A-site detector huts. C) The 
arrangement of water tanks within an A-site detector hut.}
\label{array}
\end{figure}

The Haverah Park (HP) extensive air shower array was situated at an altitude of 
220 m above sea level (mean atmospheric depth=1016 g cm$^{-2}$) at 53$^\circ$ 
58$^\prime$ N, 1$^\circ$ 38$^\prime$ W. The particle detectors of the shower 
array were water \v Cerenkov counters.  
The detectors consisted of a number of units of varying area built from water 
\v Cerenkov tank modules. The modules were of two types. The majority were 
galvanized iron tanks 2.29 m$^2$ in area, filled to a depth of 1.2 m with water 
and viewed by one photomultiplier with 100 cm$^2$ photocathode. 
A minority of 
detectors were 1 m$^2 \times $ 1.2 m deep-water \v Cerenkov detectors 
constructed from expanded plastic foam. Detector areas larger than 2.29 m$^2$ 
were achieved by grouping together a number of the larger modules in huts. 
Fig.~\ref{array}A shows the layout of the Haverah Park array.

The trigger rate of an air-shower array at large zenith angles is extremely
sensitive to the geometry of the array. Factors such as the shape and relative
altitude of detectors become very important for such showers. The relative
altitudes and orientations of the four A-site detectors, the triggering
detectors, are shown in Fig.~\ref{array}B. A gradient across the array is
apparent and this has a significant effect on the observed azimuthal
distribution. Fig.~\ref{array}C shows the positions of individual tanks within
the thermostated huts that housed the detectors. The signals from 15 of the 16
tanks, each of area 2.29~m$^{2}$, were summed to provide the signal used in
the trigger. One tank in each hut was used to provide a low gain signal. See
\cite{haverah} for a more detailed description of the array.

The signal released in a water \v Cerenkov detector is proportional to 
the energy lost in the tank by ionization. 
As most of the energy of a vertical air shower at ground level is carried by the 
electrons and photons, this technique is very effective at measuring the 
energy flow in the shower disc. Water-\v Cerenkov densities were expressed and 
recorded in terms of the mean signal from a vertical muon (1 vertical equivalent 
muon or VEM). It has been shown that this signal is equivalent to approximately 
14 photoelectrons (pe) for HP tanks \cite{evans}.
The formation of a trigger was conditional on: (i) A density of $>0.3$~ VEM~m$^{-2}$ 
in the central detector (A1) and (ii) at least 2 of the 3 remaining A-site detectors 
recording a signal of $>0.3$~VEM~m$^{-2}$. 
The rates of the triggering detectors were monitored daily. 
Over the life of the experiment, after correction for 
atmospheric pressure effects, the rates of the detectors were stable to better 
than 5$\%$. 
Approximately 8000 events with zeniths exceeding $60^\circ$  were recorded 
during an on time of $3.6~10^8$~s between 1974 and 1987. 

\subsection{Detector response}
\label{signalhor}

The calculation of the water-\v Cerenkov signal from inclined showers is 
complex. The simulation of the propagation of vertical and inclined 
electrons, gammas, and muons of different energies through Haverah Park 
tanks has been performed using a specifically designed routine 
WTANK \cite{Wtank} which uses GEANT \cite{GEANT}. 
The mean signal of electrons, gammas and muons have been convolved with the 
particle distributions obtained in the shower simulations to calculate the 
measured signal at ground by the water tanks for different zenith angles. 
Details of this calculation can be found in \cite{rate}.

The signal produced by \v Cerenkov light from the muons in the Haverah 
Park tanks is proportional to the track which typically goes through the whole tank. 
For a given muon density the signal is also proportional the tank area and as 
a result the mean signal is proportional to the tank volume and independent of 
the arrival direction of the shower relative to the tank. At large zeniths the 
smaller cross sectional area presented by the tank means that fewer muons than 
for vertical showers make up the same average signal by having longer tracks. 
Therefore Poisson fluctuations in the total number of muons going through a tank 
become more important for large zenith angles.

The signal produced by very inclined muons is enhanced by two
processes. For very inclined showers it is possible for \v Cerenkov
photons to fall directly onto the PMT without reflection from the tank walls
(we refer to such photons as ``direct light''). Also the mean muon energy
rises with zenith so that the probability of interaction in the tank is
increased because both the cross sections and the average amount of water
traversed increase.  The production of secondary electrons via pair
production, bremsstrahlung, nuclear interactions (collectively referred to as
PBN interactions), and electron knock-on ($\delta$-rays) is therefore
enhanced. For example the correction due to $\delta$-ray production increases
from 2 pe at typical vertical muon energies of 1 GeV to around 3 pe for $>10$
GeV. These contributions have been parameterized as a function of zenith and
azimuth for each of the different geometries of detectors that were used in 
the HP array.

\begin{figure}
\ybox{0.35}{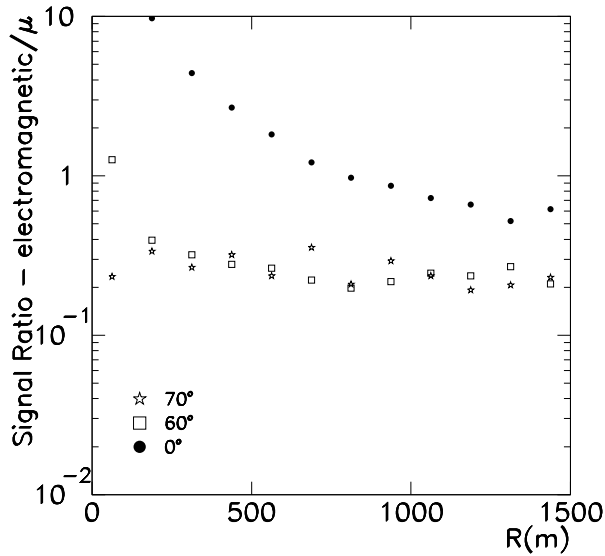}  
\caption{The ratio of the electromagnetic to muon 
contributions to water-\v Cerenkov signal as a function of distance 
from the shower axis. The non uniformity of the curves is due to statistical 
fluctuations.}
\label{shower2}
\end{figure}

On the other hand the electromagnetic particles in inclined showers usually
get completely absorbed in the tanks and the output signal is just
proportional to the input particle energy. Thus their contribution to the
total signal at larger zenith angles is suppressed compared to muons because
of the reduction of the projected area of the detectors.  In
Fig.~\ref{shower2} we show the ratio of electromagnetic to muon signal as
simulated in a \v Cerenkov tank of 1.2~m depth (as used in Haverah Park and
being implemented for the Auger Observatory) as a function of distance to
the shower axis for a vertical shower compared to two showers at large zenith
angle. The shower particles have been fed through the tank simulation as if
they were coming from the vertical direction to eliminate geometric tank
effects. The results illustrate the behaviour of the electromagnetic to muon
signal ratio because of the ratio of electromagnetic particles to muons
varying with zenith angle and distance to shower axis $r$. It is well know
that the muon lateral distribution is flatter than that for the
electromagnetic component and thus the ratio decreases below 1 for $r$ greater
than $\sim 800~$m for vertical showers. The graph illustrates that already for
zenith angles of $\sim 60^\circ$ this ratio is around the $25\%$ level for $r
> 200~$m and that for zeniths above this value this ratio at $1.5$~km is 
still about a factor 3 smaller than for vertical showers.  The rise of 
the ratio in Fig. \ref{shower2} at small distances to the core can 
be attributed to the pion showering process combined with $\pi^0$ decay.
 
\begin{figure}
\ybox{0.6}{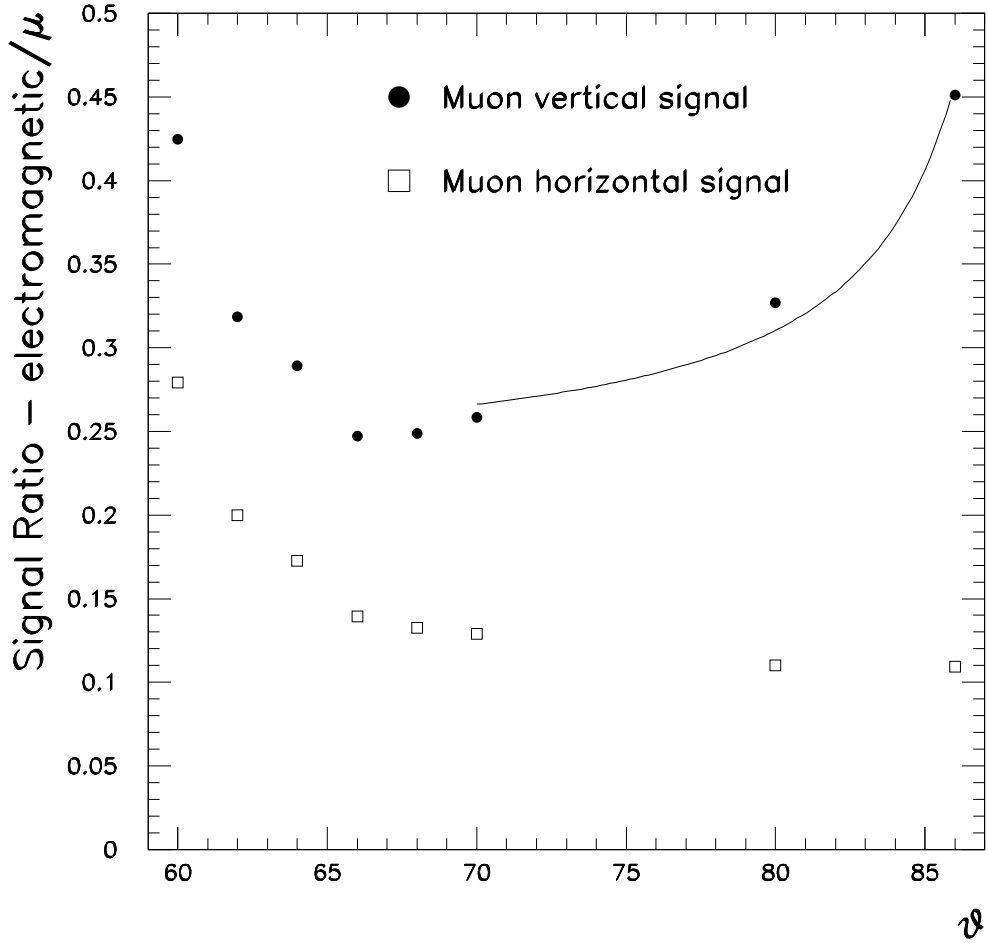}   
\caption{Ratio of the total signal from muons and electromagnetic particles
arriving within 2 km from the shower core as a function zenith angle.  The
dots corresponds to the case in which the muons are fictitiously assumed 
to enter in the tank parallel to the vertical direction and
the squares corresponds to the real signal given by a muon at the 
corresponding zenith angle (including direct light, knock-on electrons,..). 
The plot was done with 100 proton showers at $10^{19}~$ eV simulated with AIRES 
and QGSJET hadronic generator for each zenith angle. The curve shown 
corresponds 
to the fit described in the text.} 
\label{total_signal}
\end{figure}

We have averaged the water-\v Cerenkov signal induced by muons and
electromagnetic particles for $r<2~$km.  In Fig. \ref{total_signal} we plot
the average electromagnetic signal induced per muon, measured in VEM.  The
behaviour of this curve has as minimum at $\theta \simeq 67^\circ$. For zenith
angles smaller than this there is still a contribution from the
electromagnetic component from $\pi^0$ decay so the ratio is increasing
rapidly as we move towards lower zenith angles. For zenith angles above this
minimum the electromagnetic signal is dominated by muon decay which again
increases at very large zenith angles.  As the electromagnetic signal tends to
be completely absorbed in the tank, the shape of the tanks is not important for 
this figure.  We have parameterised the percentage 
contribution of the signal due to muon decay relative to the muons as a 
linear function on $\sec \theta$ independently for proton, iron
and gamma primaries, see Fig.~\ref{total_signal}.  
These relative values are useful for event simulation on
the basis of the muon density maps.

Also shown is the ratio of average signals induced by electromagnetic particles to 
that of the muons. This last curve shows how the relative contribution to the measured 
signals of electromagnetic particles decreases with zenith angle, in spite of the 
increase of the absolute electromagnetic signal per muon. This is because the muons 
from very inclined showers give enhanced signals in the tanks because of geometry. 
%

\begin{figure}
\ybox{0.6}{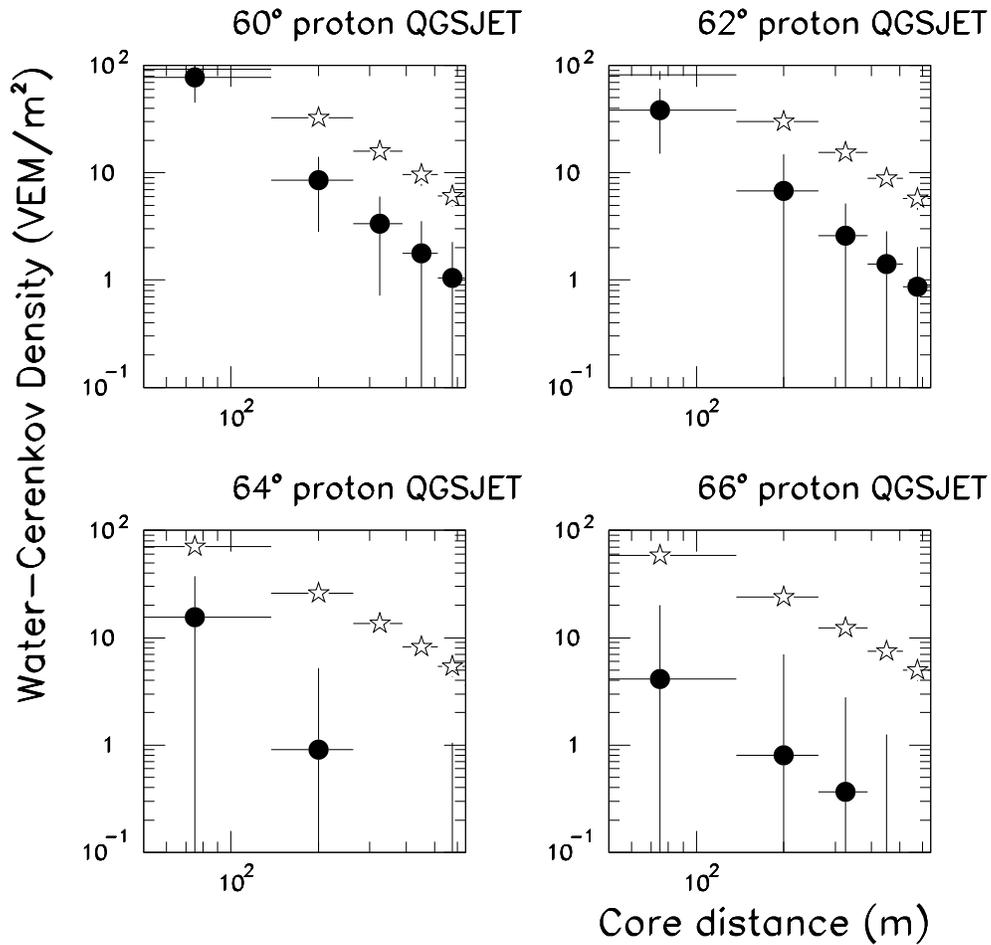}   
\caption{The mean electromagnetic signal due to the showering process and 
 $\pi^0$ decay in a HP tank (circles) compared to the muon signal (stars) 
 as a function of distance to the shower axis for four different zenith
 angles. The plot was done with 100 proton showers at $10^{19}~$ eV
 simulated with AIRES and QGSJET hadronic generator. The large errors 
 in the mean electromagnetic signals are statistical.} 
\label{emtail}
\end{figure}

After subtracting the flat component due to muon decay we have plotted in
Fig. \ref{emtail} the remaining electromagnetic contribution to the signal in
a specific HP tank configuration (the triggering tanks) for $60^\circ$,
$62^\circ$, $64^\circ$ and $66^\circ$ as a function of the distance to the
shower axis. This is the contribution from $\pi^0$ decay with large errors
because of the subtraction procedure. We have also plotted the muonic
contribution including geometric effects and enhancements due to direct light
and muon interactions.  As the zenith angle increases the electromagnetic
contribution is suppressed.  It can be seen that the electromagnetic
contribution due to $\pi^0$ decay is only relevant at small
distances($<200~m$) to the shower axis.  This contribution to the
electromagnetic component has been parameterized as a function of zenith for
proton, iron and photon primaries fitting the curves in Fig. \ref{emtail} to a
Haverah Park type \cite{HP} lateral distribution function.

\section{Event reconstruction}
\label{algo.sec}

The distortion of the circular symmetry in very inclined air showers prevents 
the use a single parameter to measure the shower energy. This is in contrast 
to near-vertical showers for which the measurement of the density at 600~m 
$\rho(600)$ has been shown to be fairly independent of composition for the 
Haverah Park array \cite{rho600}. 
Because of energy 
scaling of the muon number that controls the recorded signal at large angles, 
the natural way to obtain the energy of single events is to fit the energy and 
core position simultaneously to the expected density maps appropriate to the 
corresponding arrival direction. We describe our approach below.

\subsection{Direction reconstruction}

The Haverah Park arrival directions were determined originally using only the 
4 central triggering detectors . We have reanalyzed the arrival directions of 
showers having original values of $\theta$ $>56^\circ$, taking into account 
all detectors which have timing information. This reanalysis produces smaller 
arrival direction uncertainties due to the larger baselines involved. 

The curvature of the shower front has been investigated in section 2 using the 
AIRES code for inclined showers. The measurement error of the times recorded by 
HP array is $\sim 40$~ns, so from Fig. \ref{curvfront} it is apparent that the 
shower front is consistent with the approximation of a spherical front centred 
at the mean muon distance to production site (e.g.  at 60$^{\circ}$ the radius 
of curvature is 16 km).  Beyond $\sim 80^{\circ}$ curvature effects are rather 
small and it is quite sufficient to assume a plane front~\cite{billoir}. When 
the detected muon number is small there is a systematic effect on the curvature 
correction and large fluctuations due to limited sampling of the shower front. 
Therefore, we used only the timing information from detectors with $>15$ 
equivalent muons detected.

The direction fits of the data were originally performed using the maximum 
likelihood algorithm described in \cite{tausquared}, which is only suitable to 
fit to a plane front.  The uncertainty used in making the plane fit was:
\begin{equation}
 \Delta t({\rm ns})= \Delta t_m + {20 {\rm ns} \over \sqrt{N_\mu}} \, , 
\end{equation}
where $\Delta t_m$ is the measurement error ($\sim~40~$ ns), and the second 
term is added to account for sampling errors. 

 To fit the direction taking into account the curvature effects we first fitted 
the recorded times to a plane front. Then each measured time was corrected for 
curvature effects 
and the fit was repeated.
If the resulting zenith angle differed by more than $0.1^\circ$ from the 
previous one, times were again corrected and the fit repeated. The iteration 
was terminated when convergence had been achieved (i.e. $\delta \theta < 
0.1^\circ$).  Because of the dependence of the curvature fit on the position of 
the shower core, the iterative process must also involve fits to the particle 
density to obtain the core position. 
The implementation of this complex iterative procedure will be 
described in a subsection \ref{densityfit}. 
The uncertainty expression used in the  
curvature fit is:
\begin{equation}
 \Delta t({\rm ns})=\sqrt {\Delta t_m^2 +\Delta t_c^2 +\Delta t_s^2} \, ,
\end{equation}
where $\Delta t_c$ is the error induced in the corrected times by the
uncertainty in the core position, $\Delta t_m$ is the measurement error, 
and $\Delta t_s$ is the sampling error (see subsection \ref{tonto} ).

\subsection{Parameterizations for the muon densities}

We have obtained the inputs needed in our analytical description of muon 
densities \cite{model} from specific AIRES simulations with the 
QGSJET \cite{QGSJET} hadronic interaction generator.  
For three possible compositions of  proton, iron and gamma primaries,  
one hundred showers were generated for each zenith angle in the range 
$60^\circ$~-~$89^\circ$ (in $1^\circ$ steps) in the absence of a magnetic field, 
at a fixed energy of $10^{19}$ eV. 
Using the procedure described in \cite{model}, 
we have prepared a compact library of muon density patterns at a 
fixed energy for different zenith angles and different compositions. 
Magnetic deviations are accounted for in 
the muon distributions projecting the Earth's magnetic field onto the 
transverse plane and using the algorithm described in \cite{model} which 
rotates the pattern depending on the azimuthal direction. 
Different energies were obtained through energy scaling as indicated in 
subsection \ref{muonshowers}.

The electromagnetic component is separated into two parts:

$\bullet$ {\it The component fed by muon decay}: In the previous section we 
showed that the contribution to the signal in the tanks due to electromagnetic 
particles produced by muon decay was present at all core distances and that it 
made a contribution to the signal that depends slightly on zenith angle and 
is of $\sim 3$ photoelectrons per arriving muon. 
The spatial distributions of this electromagnetic contribution follows 
the muon density pattern, so it is relatively simple to include it using the 
density maps described above. 

$\bullet$ {\it The component fed by $\pi^0$ decay}: The tail of the
electromagnetic part of the shower contributes mildly to the particle density
at ground level at zenith angles below $70^{\circ}$ and core distances less
than 500~m.  This contribution has been modeled using AIRES with QGSJET (see
previous section) and is radially symmetric in the transverse plane. The tail
of the electromagnetic part of the shower contributes 20 \% of the total
water-\v Cerenkov signal at 400~m from the core for a $60^{\circ}$ shower. As
is clear from Fig.~\ref{emtail}, the contribution drops both for larger
distances and for higher zenith angles. This electromagnetic component was
calculated at an energy of $10^{19}~$ eV: the values for different energies
were obtained by scaling with energy ($\rho_{em}\propto E_0$).

\subsection{Detector signal conversion}

We will later on compare the signal at the detectors to predictions based on 
simulation of showers. 
For each detector we will compare the {\sl recorded} number of muons $N_\mu^r$ to 
the number of muons {\sl predicted} from a given density map $N_\mu^p$, which is 
simply obtained multiplying the muon number density by the transverse area for each 
detector. The actual values of $N_\mu^p$ used are corrected to account 
for the electromagnetic contribution due to the tail of the showering 
processes. We now describe the process of converting the actual recorded signal 
to $N_\mu^r$ which is not straight forward because of several corrections 
that need to be considered. 

The detector signals were recorded in units of vertical equivalent muons. Using 
the GEANT based package, WTANK \cite{Wtank}, we have found that this unit 
corresponds to an average number of 14 photoelectrons, in agreement with 
experimental estimates \cite{evans}. For inclined showers additional effects, 
such as direct light on the photomultiplier tubes, delta rays, pair 
production and bremsstrahlung by muons inside the tank, increase this number.  
For a given zenith angle, we have calculated the mean number of photoelectrons 
per muon ($pe_\mu$) taking into account all the processes mentioned before 
except for pair production and bremsstrahlung. 
Pair production and bremsstrahlung do not alter the expected rate 
as a function of zenith angle by more than a $1\%$, so we have not included this 
effect to save computing time. 

To calculate the value of $pe_\mu$ we use:
\begin{equation}
pe_\mu = (pe_{vem}+pe_{\delta}) \frac{A_v}{A_h(\theta,\phi)}+pe_{em}(\theta) 
+pe_{dl}(\theta) \, ,
\label{pemuon}
\end{equation}
where $pe_{dl}$ is the contribution from the direct light, $pe_{em}$ is the 
contribution of the electromagnetic part from muon decay which is $\sim 3$ 
photoelectrons per arriving muon. The first term is the contribution 
proportional to the muon track, including the \v Cerenkov light from both 
the muon track ($pe_{vem}$) and from the $\delta$ rays 
($pe_{\delta}$), which have to be corrected by 
the ratio of the vertical to the inclined average tracklength. This correction 
can be also expressed in terms of the ratio of the cross sectional areas presented 
by the tanks for vertical and inclined muons ($A_v / A_h$) as explained in 
subsection \ref{signalhor}. 

The different sizes of detectors present in Haverah Park array are described in 
Table \ref{tab2} with their corresponding areas, density thresholds and 
saturation densities. These differences have forced us to simulate with WTANK  
the different detector geometries for different zenith angles to obtain the 
corresponding values of $pe_\mu$.
\begin{table}
\begin{center}
\begin{tabular}{|lccc|} \hline
Type & Vert. area ($m^{2}$) & Thresh. (VEM $m^{-2}$) & Sat.(VEM $m^{-2}$)  \\ \hline \hline
Trigger detectors & 37. & 0. & 45. \\
2 km array &  14. & 0. & 45. \\
150 m array & 9. & 0. & 60.\\
Infill array & 1. &  7. & - \\
J,K,L detectors & 2.25 & 7. & - \\
\hline
\end{tabular}
\end{center}
\caption{Characteristics of the different kind of detectors in the Haverah Park array}
\label{tab2}
\end{table}

The recorded signals at each detector are first converted into the corresponding  
number of photoelectrons by multiplying the recorded density (m$^{-2}$) by the 
vertical area and the number of photoelectrons per vertical muon (14 pe). The 
number of muons going through each tank, $N_\mu^r$, is then obtained dividing this 
number of photoelectrons by the number expected per muon at the corresponding 
arrival direction $pe_\mu$, given in Eq. (\ref{pemuon}). 
For detectors that saturate or have thresholds, the corresponding number of 
muons $N_\mu^{sat}$ and $N_\mu^{th}$ are calculated for a given arrival direction 
in an analogous fashion using the saturation and threshold signals of Table \ref{tab2}. 

\subsection{The fitting algorithm}
\label{densityfit}

The observed densities were fitted against predictions using the maximum 
likelihood method. The quantity to maximize in this method is:
\begin{equation}
\ln ~P(x_c,y_c,E_0)=\ln (P_1~P_2~....~P_n)=\sum_{i=1}^n~\ln~P_i \, ,
\label{likehood}
\end{equation}
where $n$ is the number of detectors used in the fit and $P_i$ is the 
probability that the $i^{th}$ detector records $N_\mu^r$ muons if 
the predicted number of muons is $N_\mu^p$ (as obtained from the muon density 
maps). The primary energy $E_0$ and the core coordinates ($x_c,y_c$) are the 
free parameters in the fits. 
In order to calculate the probabilities needed in Eq.~\ref{likehood} we 
assume a Poisson distribution with mean $N_\mu^p$ given by:
\begin{equation}
P_i=\frac{\left (N_\mu^p \right )^r e^{-N_\mu^p}}{r!} \, ,
\label{poisson} 
\end{equation}
where $r$ is the closest integer to $N_\mu^r$. When large numbers of 
muons ($N_\mu^r>8$) are involved we approximate the Poisson 
distribution with a gaussian distribution with mean $N_\mu^p$ and width $\sigma$ 
obtained adding three different errors in quadrature:
\begin{equation}
\sigma=\sqrt{\sigma_p^2 + \sigma_m^2 +\sigma_g^2  }  \, ,
\label{sigma}
\end{equation}
where $\sigma_p=\sqrt{N_\mu^p}$ is the Poisson error of the muon number, 
$\sigma_m$ is the measurement error ($7\%$ of the recorded signal), and 
$\sigma_g$ is the error induced by geometrical considerations: dependence of the 
detector area with azimuth and azimuthal variations of the direct light. The 
main contribution to $\sigma$ comes from $\sigma_p$.
If the detector is saturated the corresponding 
probability is calculated integrating the gaussian distribution 
from $N_\mu^{sat}$ to $\infty$. If the detector 
density is under the threshold we evaluate the Poisson probability of getting 
$N_\mu^{th}$ or fewer muons. 

A three-dimensional grid search was made to maximize Eq.~\ref{likehood} finding  the most likely impact point and shower energy. The energy was varied in 
the range $10^{17} < E_0^p < 10^{21}$~eV in 
steps of $0.1$ in $\log_{10}(E_p/\mbox{eV})$. The impact point was varied over a 
grid of 12 km $\times$ 6 km in 40 m steps in the perpendicular plane, the grid 
asymmetry being necessary to accommodate the ellipticity of inclined showers.  

Since angle reconstruction depends on the core position for curvature 
corrections a complex algorithm was required to avoid spurious dependences 
between core location and direction determination.  The steps of the algorithm 
to find the final parameters of an event are the following:

$\bullet$ {\it 1-} Find $\theta$, and $\phi$ by fitting a plane front to the times 
registered by the triggering detectors.

$\bullet$ {\it 2-} With the reconstructed direction, find the core position and 
primary energy through a three dimensional grid search, maximizing the 
likelihood function.

$\bullet$ {\it 3-} Find a new value for $\theta$, and $\phi$ fitting a plane front 
to the times registered by the detectors within 1 km of the core (found in 
the previous step) in the shower plane.  If there are less than 7 detectors with time 
information we complete the number with the next nearest detectors, which may 
lie $>1$ km from the shower axis.

$\bullet$ {\it 4-} With the reconstructed direction, find a new core position 
and primary energy.

$\bullet$ {\it 5-} Repeat step 3 and 4 once to avoid any bias induced by the 
first determination of the shower core (which used a direction fitted with a 
small number of times from detectors that could be far away from the shower 
core).

$\bullet$ {\it 6-} Find $\theta$ and $\phi$ taking into account the curvature
in the front. This yields the final reconstructed direction. We also calculate
$\Delta \theta$. The zenith angle does not usually change more than $1^\circ$
compared with the value obtained in the previous step.

$\bullet$ {\it 7-} With the reconstructed direction, find again the core 
position and primary energy ($x_c,y_c,E_0$). This will be the final 
reconstructed parameters of the event.

$\bullet$ {\it 8-} Find core position and primary energy for changes to the 
value of $\theta$ by $\theta+\Delta \theta$ and $\theta-\Delta \theta$. This 
step is particularly important for controlling and understanding the systematic 
uncertainty of the primary energy due to the zenith angle uncertainty.

Errors in the energy and core determination were determined from the likelihood 
function as described in \cite{lampton}.  In addition to this error, an error in 
energy arises due to the uncertainty in the zenith angle. The error from the 
zenith angle determination and the error from the fit are added in quadrature to 
give the total error shown in Table \ref{tab}.  The typical error 
in the position of the core is $100$~m and in $\log_{10}E_0$ it is $0.1$, 
corresponding to $26 \%$.  

\subsection{Results of the data fit}

\label{res.sec}

Over 8000 events were fitted with muon density maps generated for proton 
primaries and the QGSJET hadronic generator, following the procedure explained 
in the previous subsection. 

To guarantee the quality of events the following cuts were made to the 
reconstructed events: 
(i) the distance from the central triggering detector to 
the core position in the shower plane is required to be below $r_{max}=2$~km, 
(ii) the $\chi^2$ probability for the energy and direction fits must be $>$ 1 \%, 
(iii) the downward error in the energy 
determination is required to be less than a factor of 2.  
The chosen value of $r_{max}$ guarantees that  
the core position is always surrounded by detectors in the HP array.  
After making the cuts described above we found 52 events with $E_0 > 10^{19}$ 
eV, ten events with energies above $4 \times 10^{19}$~eV and one with energy 
above $10^{20}$ eV. 
For zenith angles greater than $80^{\circ}$ no showers pass cut (iii). 

\begin{figure}      
\twobox{0.32} {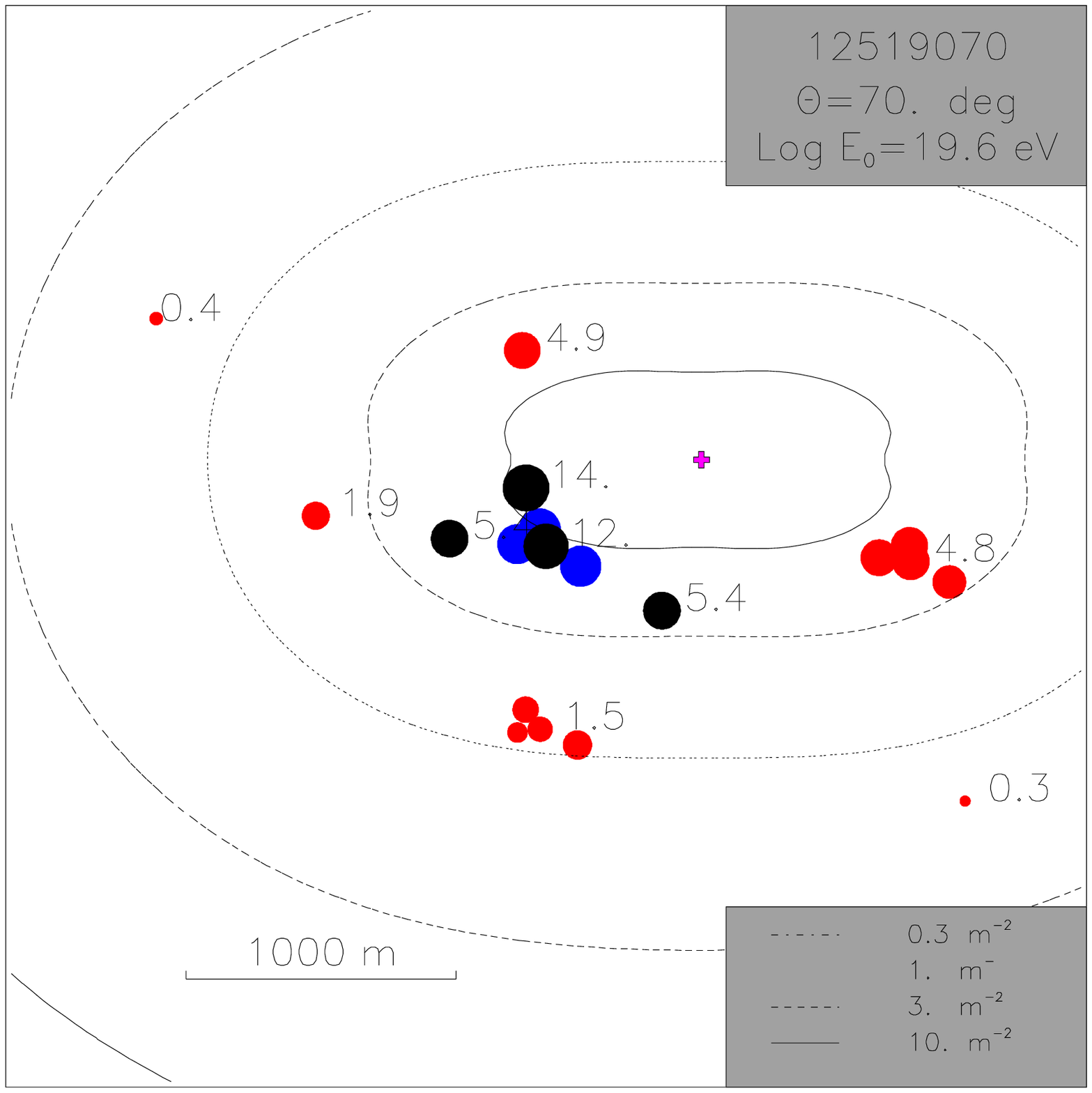} {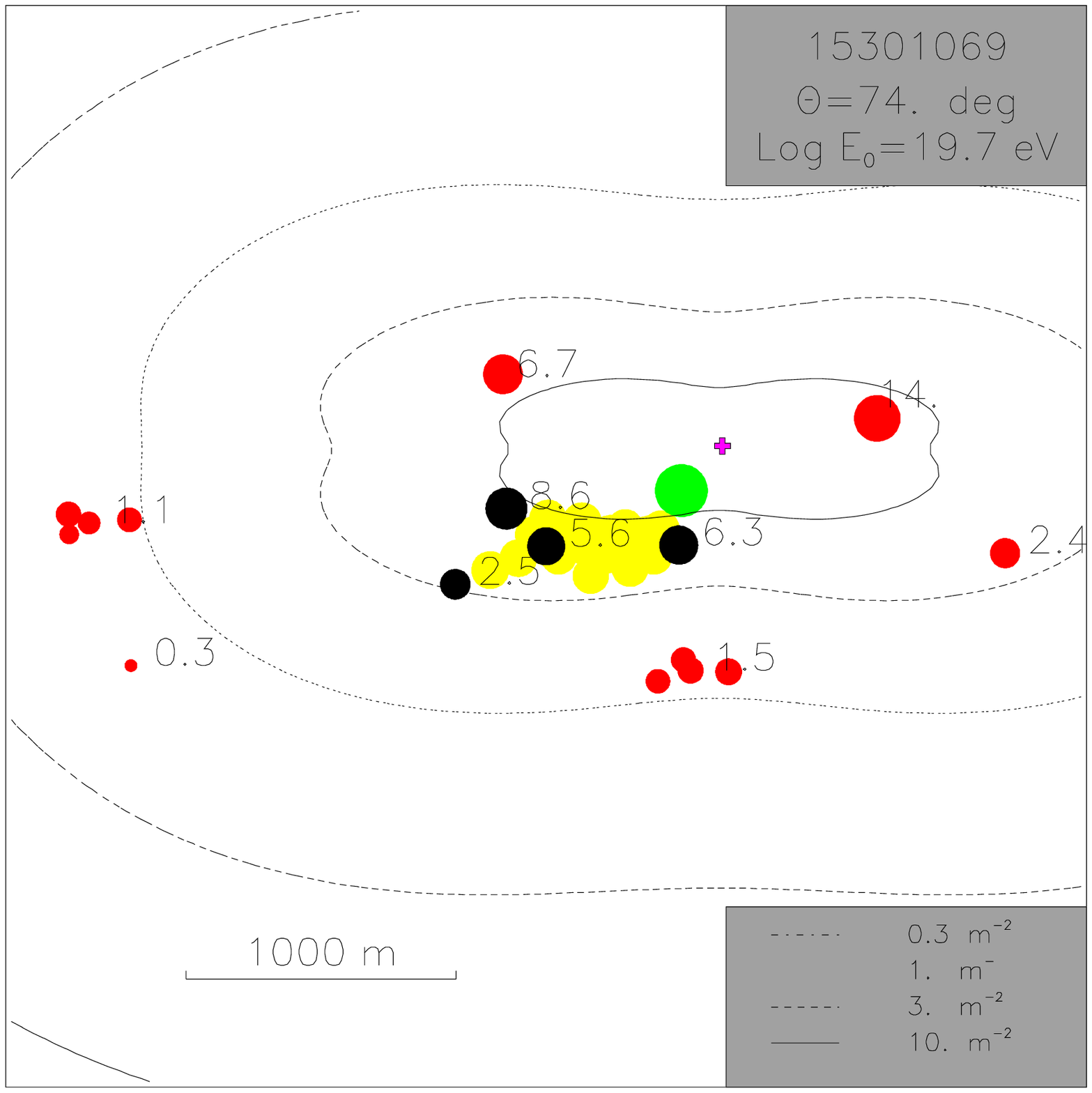}
\caption{Density maps of two events in the plane perpendicular to the shower
axis. Recorded muon densities are shown as circles with radius proportional to
the logarithm of the density. The detector areas are indicated by shading; the
area increases from light grey to black as 1, 2.3, 9, 13, 34 m$^{2}$. 
The position
of the best-fit core is indicated by a cross. Selected densities are also
marked. The $y$-axis is aligned with the component of the magnetic field
perpendicular to the shower axis.} 
\label{event1}
\end{figure}


\begin{figure}      
\twobox{0.32} {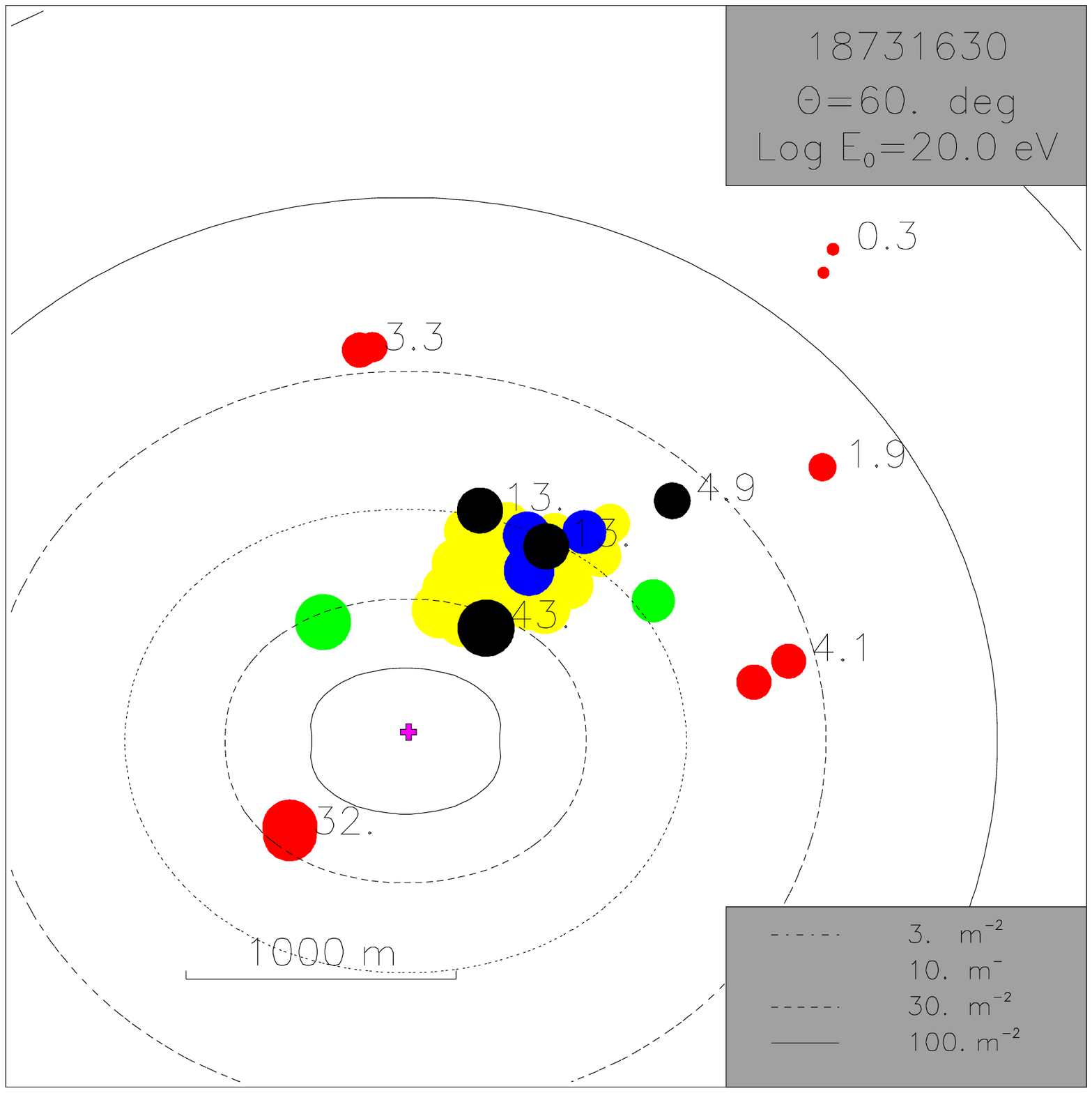} {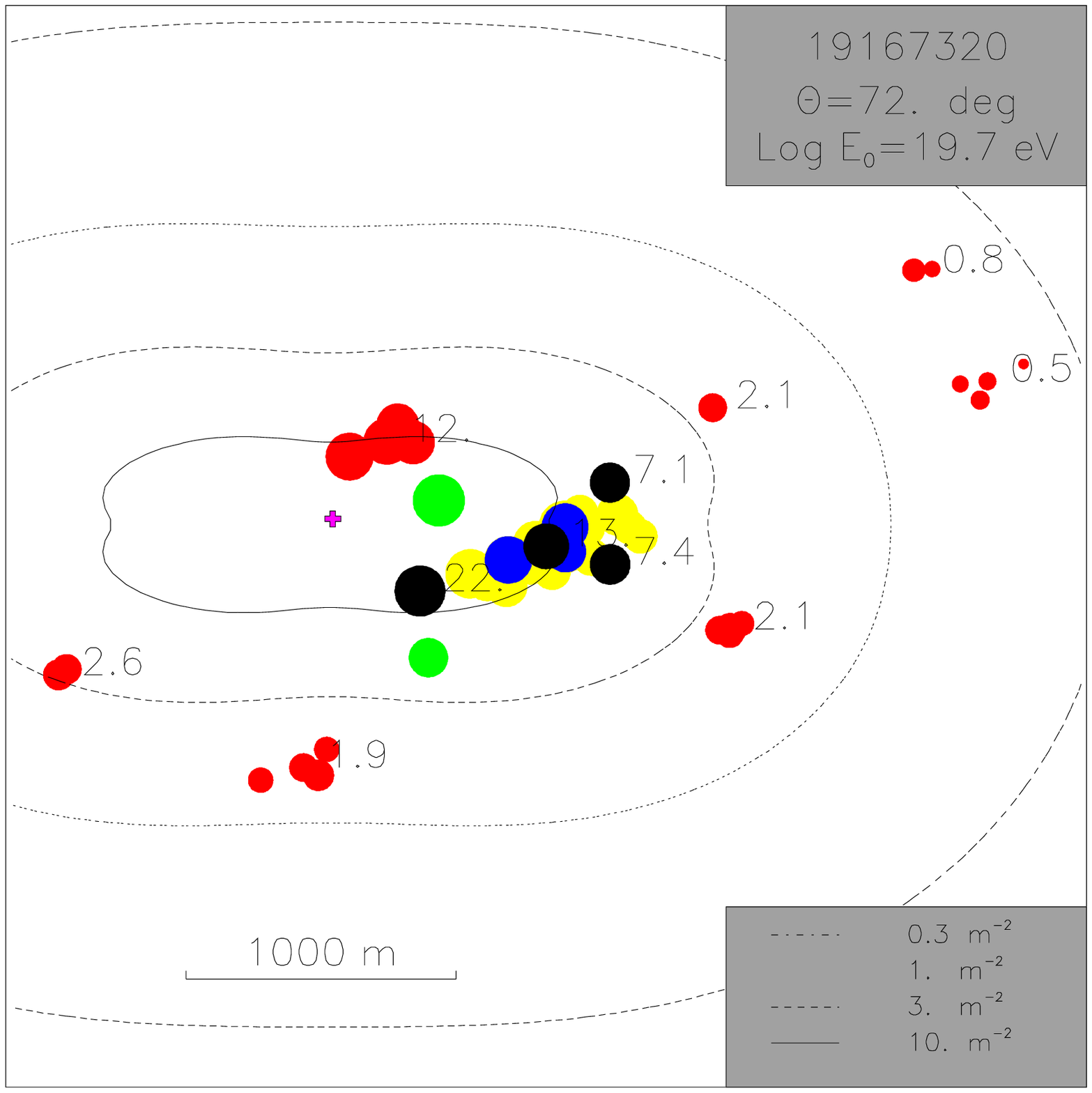}
\caption{Same as Fig. \ref{event1}}
\label{event3}
\end{figure}


In Figs. \ref{event1} to \ref{event3} the density maps for four reconstructed
events are shown in detail. These maps are plotted in the plane perpendicular
to the shower direction together with the contours of densities that best fit
the data. In each figure the array is rotated in the shower plane such that
the $y$-axis is aligned with the component of the magnetic field perpendicular
to the shower axis. In Fig.~\ref{event1} and on the right panel 
of Fig~\ref{event3}
the asymmetry in the density pattern due to the geomagnetic field is
apparent. For all these events the core is surrounded by recorded densities
and is well determined.  In table \ref{tab} details are given of the 10 events
with $E_{p} > 4 \times 10^{19}$ eV.

\begin{table}
\begin{center}
\begin{tabular}{|lcccccccc|} \hline
MR & \multicolumn{2}{c}{ Zenith ($^{\circ}$)} & RA ($^{\circ}$) & Dec. ($^{\circ}$) & 
\multicolumn{3}{c}{$\log_{10}(E_{p}/\mbox{eV})$} & $\chi^{2}/\nu$ \\ \hline \hline
18731630 & 60 &$\pm$2.3 & 318.3 & 3.0  & 20.04 & -0.03 & +0.03 & 40.0/42  \\
14050050 & 65 &$\pm$1.2 & 86.7  & 31.7 & 19.89 & -0.08 & +0.10 & 11.0/13  \\
18565932 & 68 &$\pm$1.3 & 46.4  & 6.0  & 19.88 & -0.22 & +0.34 & 15.5/15 \\ 
25174538 & 65 &$\pm$1.2 & 252.7 & 60.2 & 19.85 & -0.22 & +0.20 & 5.0/5  \\
14182627 & 70 &$\pm$1.3 & 121.2 & 8.0  & 19.76 & -0.05 & +0.05 &  5.0/10 \\
15301069 & 74 &$\pm$1.2 & 50.0  & 49.4 & 19.76 & -0.06 & +0.05 & 27.1/32 \\
19167320 & 72 &$\pm$1.3 & 152.5 & 25.9 & 19.75 & -0.06 & +0.04 & 36.5/33 \\
12753623 & 74 &$\pm$2.1 & 304.9 & 17.1 & 19.67 & -0.07 & +0.10 & 11.4/11 \\
24503624 & 69 &$\pm$2.1 & 16.9  & 53.0 & 19.63 & -0.22 & +0.33 & 11.0/9  \\
12519070 & 70 &$\pm$1.3 & 47.7  & 8.8  & 19.62 & -0.08 & +0.06 & 15.2/14 \\ 
\hline
\end{tabular}
\end{center}
\caption{Zenith angle, arrival direction coordinates and shower energy 
(assuming proton primary) of selected showers with energy 
$> 4\times10^{19}$~eV. MR is the event record number. 
The reported $\chi^{2}$ values and the degrees of freedom ($\nu$) refer to 
the density fits.}
\label{tab}
\end{table}

This work improves and extends the results presented in \cite{PRL} and is
compatible with it. There are however slight differences which are due to the
improvements, namely:

 \begin{enumerate}

\item Improved muon density parameterizations, now in $1^\circ$ steps. 

\item Inclusion of densities below threshold in the fitting algorithm.

\item Better treatment of the electromagnetic part of the shower from muon 
decay.

\item Inclusion of events with original zenith angle $56^\circ < \theta < 60^\circ$.

 \end{enumerate}

The inclusion of three additional events in table \ref{tab} compared with what was 
obtained in \cite{PRL}, and the changes in the energies of some events should be 
noted. It must be stressed that the new energy always lies within the error 
quoted in \cite{PRL}, and the three new events were not included in 
the original list because they failed to pass the cut on the downwards error. 

The photoelectron distributions in a water detector show long tails due to the
processes mentioned in section \ref{signalhor}.  We therefore expect an excess
of upward fluctuations over downward fluctuations from the average detector
signal.  For each event we calculate the probability that each of the
detectors involved has a signal which deviates by more than $>$2.5$~\sigma$
from the average using the simulated photoelectron distributions. We reject
signals having (upward or downward) deviations greater than 2.5~$\sigma$,
recalculating the best-fit core after any rejection.  Of 226 densities in the
events described below and listed in table~\ref{tab} we have rejected 12
upward deviations (the expected number was 17) and a single downward
deviation.  We consider this to be a strong vindication of our understanding
of the signal in the tanks and of our modeling procedures.

\section{Generation of artificial events}

Besides the fitting of the individual events it is extremely important to
compare the data obtained with expectations.  We have simulated ``artificial''
events assuming a given energy spectrum for the cosmic rays, taken from other 
experiments and assuming different primary compositions.  In order to compare 
the simulated results to those obtained from the data, 
we must also calculate the reconstruction efficiency which is sensitive 
to the cuts made. 
Throughout we use the QGSJET as the hadronic generator of the simulations.

We have generated showers in the range of energies $10^{18}$~eV to 
$10^{21}$~eV in bins of 0.05 in $\log_{10} E_0$. For each of these energy bins we 
have adjusted the number of artificial events generated to approximately obtain  
$300$ showers that trigger the array. The procedure for generating each artificial 
event is the following:

\begin{enumerate}
 
\item We randomly select an arrival direction assuming isotropy according to 
a $\sin \theta$ distribution for  zenith angle and a uniform distribution for 
azimuth ($\phi$). 

\item Each shower is directed on to the array with a random impact point position 
in the transverse plane up to 2.5 km away from the centre of the array.

\item Each time a shower is directed at the array, the total muon 
number($N_{\mu}$) is fluctuated to take into account shower fluctuations. For 
proton and iron primaries we used a gaussian distribution of spread 
0.2$~N_{\mu}$, and for photon primaries we used the distribution in Fig. 
\ref{fluct_ph}.

\item The density in the ground plane at the location of each of the detectors 
is read from the library of muon density maps.

\item The corresponding signal and arrival time in each of the detectors is 
generated (see next subsection).

\item The trigger condition of the Haverah Park array is tested.

\item If an event is deemed to trigger the array then the density and time 
information is recorded in the same format as the  real data.

\item Each artificial event is assigned a weight ($w_o$) which is 
$N_{exp}/N$, where $N$ is the total number of events generated in a 
particular energy bin, fulfilling, or not, the triggering condition and $N_{exp}$ is 
the total number of CR expected from the assumed flux for the same energy bin 
integrated over the zenith angle range considered ($59^\circ-89^\circ$). 

\end{enumerate}

The artificial events, recorded in the same format as real data, are analysed
with the same algorithm assuming a proton composition for the maps and with
QGSJET as the hadronic generator, and applying the same cuts.  The resulting
spectrum is obtained by adding the weights of the individual artificial events at
the corresponding reconstructed energies.

\subsection{Implementation of the signal in the detectors}

The signal in each detector is artificially generated as follows:

\begin{enumerate}

\item The projected area of the detector in the shower plane is calculated.

\item Given the local muon density and the projected area, the number of       
incident muons is sampled from a Poisson distribution.

\item The track length of each muon through the detector is sampled from a      
distribution obtained analytically from the detector geometry (see 
Fig.~\ref{array}C).\footnote{This distribution accounts for the       
possibility that at large zenith angles a single muon may traverse several tanks.} 

\item The contribution of indirect \v Cerenkov light from the incident muons       
and from $\delta$-ray electrons is calculated from the sampled track      
lengths (12 pe for each 1.2 m of track, with an additional 3 pe/1.2 m to account 
for the signal from $\delta$-rays).

\item The signal from direct light on the PMTs is related to the detector 
geometry in a more complex way and is implemented using WTANK to simulate the      
passage of muons through the whole detector for a range of zenith and azimuth 
angles.

\item The electromagnetic component of the shower due to muon decay is      
approximated by the addition of a number of photoelectrons per muon ($n$)  
which depends smoothly on zenith angle.

\item The electromagnetic component of the shower from $\pi^0$ decay is 
calculated using the parameterizations discussed in section 3.

\item The signal generated in this way is fluctuated according to 
measurement  errors.  

\end{enumerate}

The arrival time of the first muon is generated assuming a spherical 
shower front with radius 
equal to the mean distance to the production site of the muons at each particular 
zenith angle. The time is then fluctuated according to measurement and sampling 
errors. 

\subsection{Comparison of data and artificial event distributions}

\begin{figure}
\ybox{0.5}{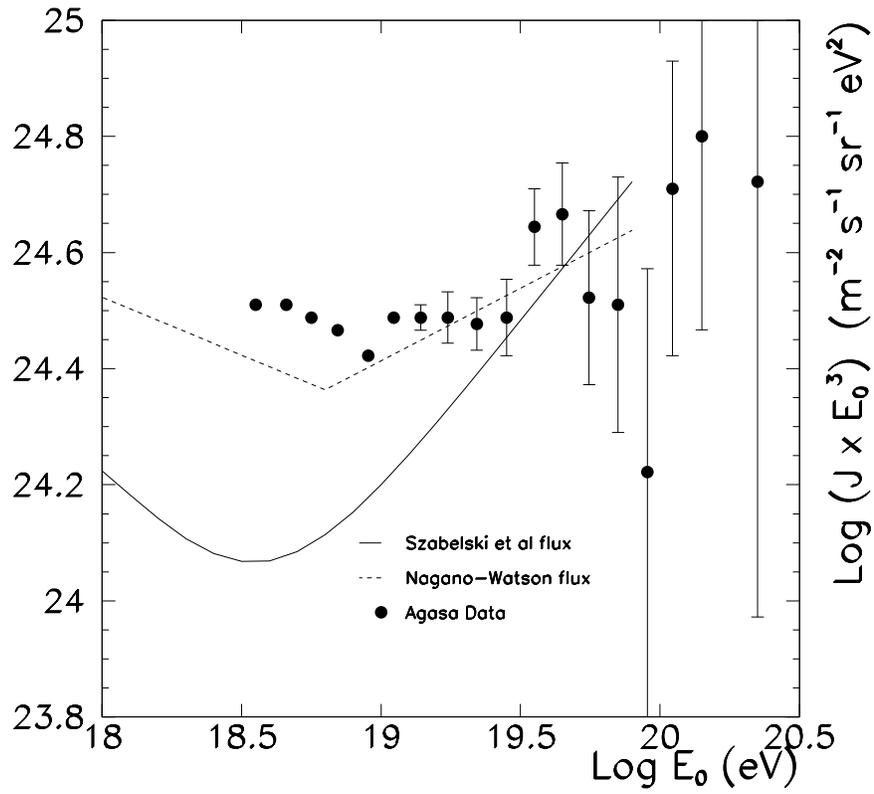}     
\caption{Parametrizations of the cosmic ray flux 
 between 10$^{18}$ and 10$^{20}$ eV used in this work 
 due to Nagano-Watson \cite{WatsonNagano} (dashed line) and to 
 Szabelski et al. \cite{Wolfendale} (full line) compared to AGASA data 
\cite{agasadata}. }
\label{arnold}
\end{figure}

We assume a recent parameterization of the energy spectrum given in
\cite{WatsonNagano} noting that the agreement between the fluorescence
estimates of the spectrum and those made by other methods implies that we have
an approximately mass independent knowledge of the spectrum measured in the near-vertical
direction. The flux above 10$^{19}$ eV is assumed to be known to within 20$\%$
uncertainty.  We will compare to the results obtained using an alternative
energy spectrum given in \cite{Wolfendale}. Both fluxes are compared in
Fig.~\ref{arnold}.

\begin{figure}
\ybox{0.3}{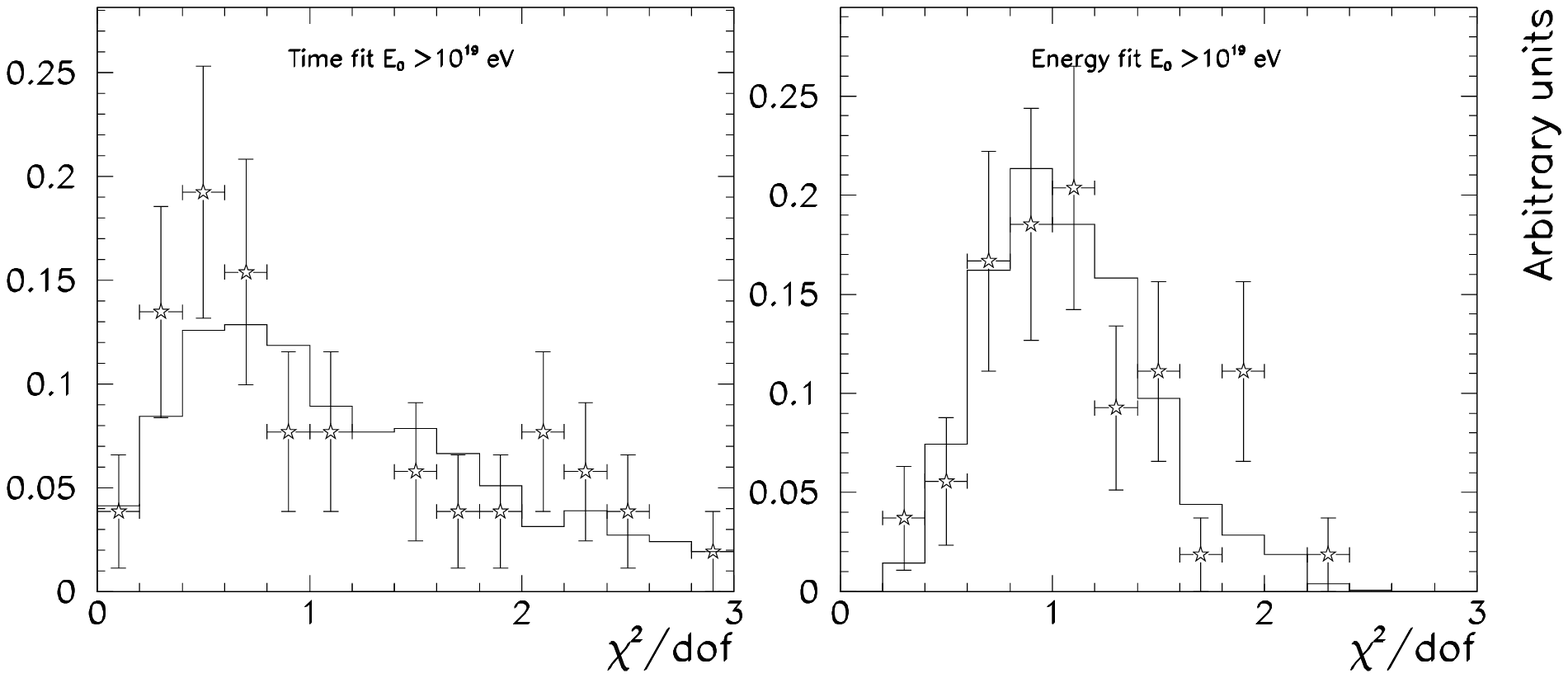}  
\caption{$\chi^2$ distributions from the energy and direction reconstruction
of data (stars) and artificial events (histogram), assuming proton 
composition and the parameterizations of the spectrum given 
in \cite{WatsonNagano}. 
}
\label{chis}
\end{figure}

\begin{figure}
\ybox{0.3}{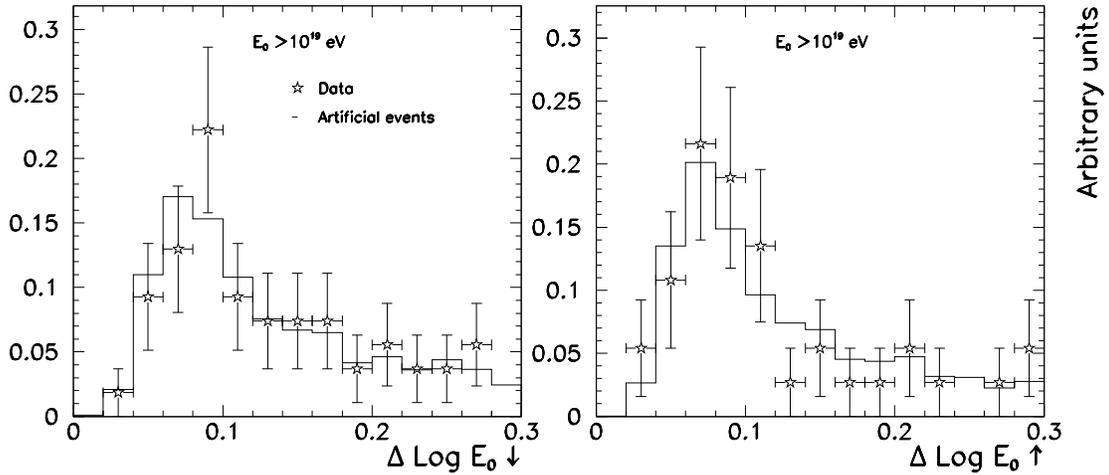}  
\caption{Downward and upward error distribution in the reconstructed 
energy from the density fits to the data (stars) and to the artificial 
events (histogram). 
}
\label{errors}
\end{figure}

\begin{figure}
\ybox{0.3}{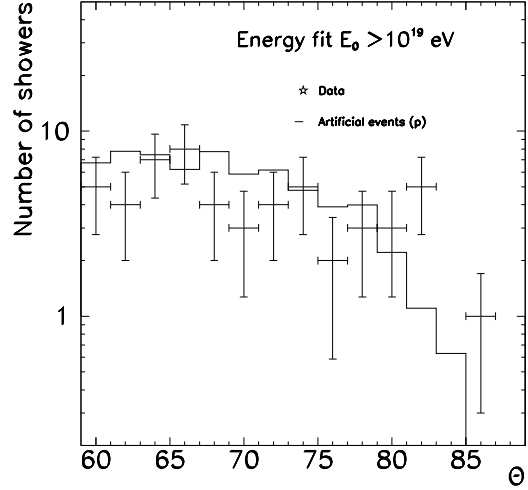}  
\caption{Zenith angle distribution for data (stars) and artificial events
(histogram). No normalization has been made. Statistical error bars are also shown.}
\label{zendist}
\end{figure}

\begin{figure}
\ybox{0.3}{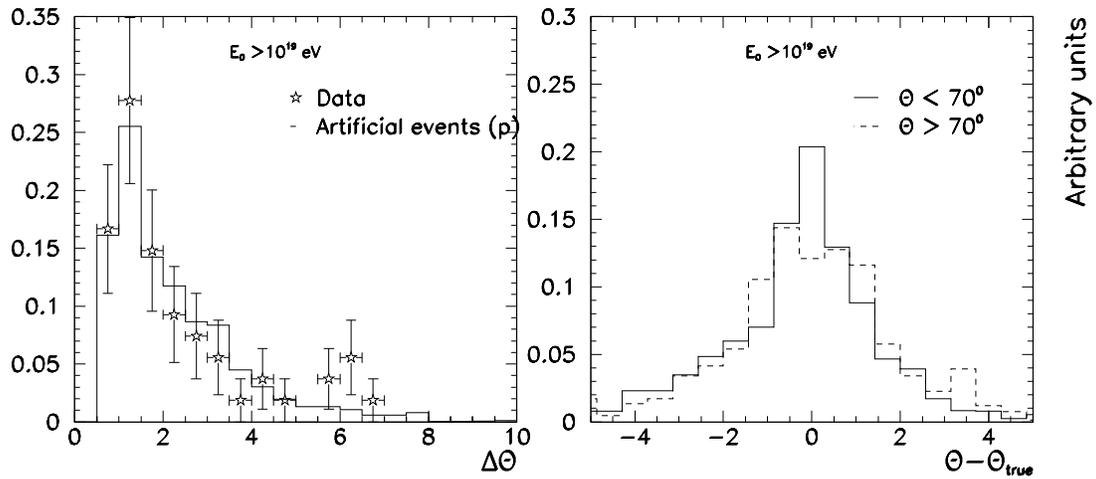}  
\caption{Left Panel: distribution of errors in zenith angle from the fit for the 
data (stars) and artificial events (histogram). 
Right panel: distribution of the difference 
between the real and the reconstructed zenith angle.} 
\label{zenerr}
\end{figure}

In Fig. \ref{chis}, \ref{errors}, \ref{zendist}, \ref{zenerr} we show 
different output parameters of the 
event reconstruction for the artificial events assuming a proton composition 
and the spectrum given in \cite{WatsonNagano}, compared to data. 
All the events used in these figures pass the cuts described in the 
previous section, in particular for energies above $10^{19}$~eV, 
$r_{max}=2~$km. 
The agreement obtained is encouraging and suggests that the simulation accurately 
mimics the data.

\begin{figure}
\ybox{0.3}{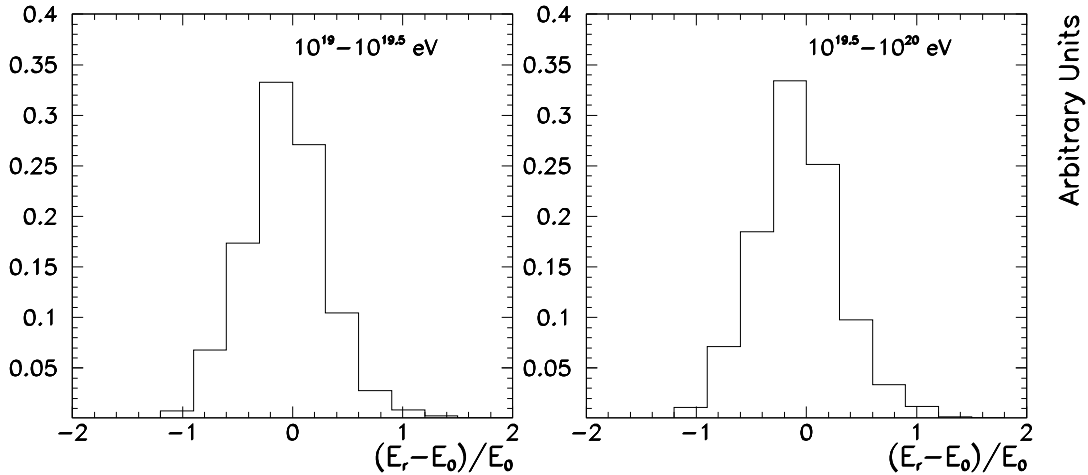}    
\caption{Energy resolution integrated for all zeniths in different energy 
bins. A flat energy distribution is assumed for each graph.}
\label{energycomp}
\end{figure}

In Fig. \ref{energycomp} we show the energy resolution for different energy
ranges.  A finite energy resolution has the effect of increasing the measured
rate by misinterpreting more abundant lower energy events as having a higher
energy.  However no corrections need be made in this approach because the same
effect is present both in data and simulations.

\newpage

\section{Limits on composition}

After all the quality cuts are implemented as discussed in section \ref{algo.sec} 
we calculate the event rate as a function of the primary energy integrating 
over all zenith and azimuth angles. 
We will concentrate here on the events with reconstructed proton energy above $10^{19}~$eV, which provide the most stringent conclusions about UHECR composition. 

\begin{figure}
\ybox{0.3}{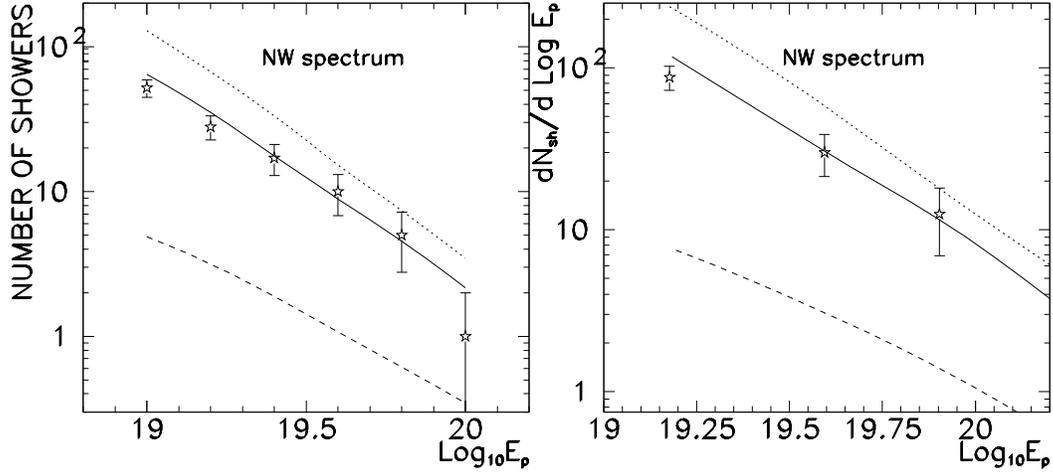}    
\caption{Integral (left panel) and differential (right panel) 
number of inclined events as a function of energy for the Haverah Park 
data set (*stars) compared to the predictions for iron (dotted line), 
protons (continuous) and photon primaries (dashed). The parameterization 
of the spectrum given in \cite{WatsonNagano} is assumed.}
\label{rate2}
\end{figure}

\begin{figure}
\ybox{0.3}{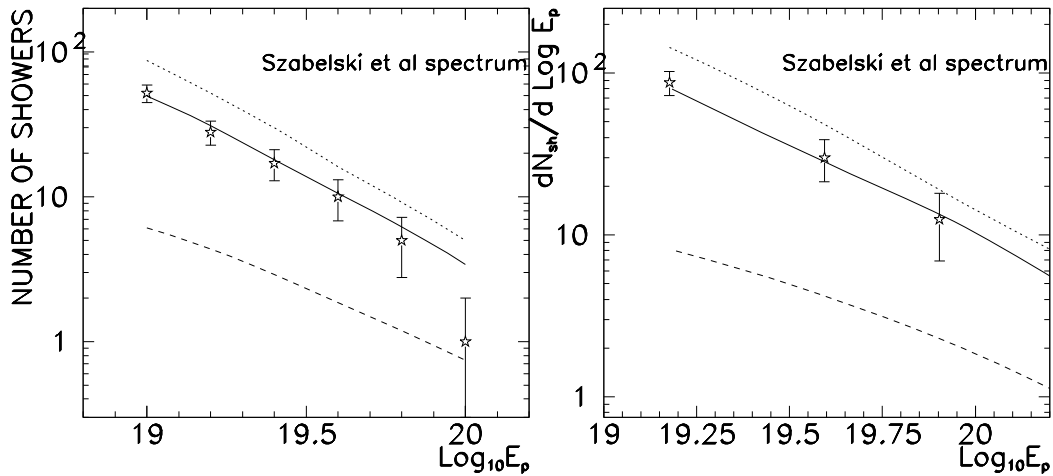}    
\caption{Integral (left) and differential (right) number of inclined 
events as a function of energy 
for the Haverah Park data set compared to the predictions for iron, 
protons and photon primaries. The parameterization of the spectrum given in
\cite{Wolfendale} is assumed.}
\label{rate}
\end{figure}

In Fig. \ref{rate2} we show both the integral and differential energy spectra
obtained from the artificial events under three different assumptions for the
primary composition (protons, iron and photons) compared to the data using the
cosmic ray parameterization given in \cite{WatsonNagano}.  We also show the
spectra obtained using the cosmic ray flux spectrum from \cite{Wolfendale},
see Fig. \ref{rate}. All curves are for the QGSJET hadronic interaction
model.  The agreement between the curves generated for protons with the two
spectra is remarkable. The normalization of the curves has not been manually
adjusted.  The expected rate increases if iron is assumed and decreases if
photons are assumed. This just a matter of counting muons, heavier nuclei have
more muons while photons are known to have much fewer muons. For the same
reason shifts in the curves can be expected if different hadronic interaction 
models are used according to the number of muons they predict.

The remarkable point about this graph is that the expected rate for
photons is about an order of magnitude below the proton prediction.
Assuming that cosmic rays have a proton/photon mixture at ultra high
energies it is easy to obtain a bound from this graph. 
We can get bounds on photon abundance at a given confidence level 
comparing the measured number of events above a given threshold and its 
error to the expected numbers in the case of proton and photon compositions, 
taking into account the uncertainty in the prediction from the 
normalization error in the parameterization of the cosmic ray spectrum. 
Assuming the 
prediction obtained with the flux in \cite{WatsonNagano} we obtain
that less than $48\%$ of the observed events above $10^{19}~$eV can be
photons with a $95 \%$ confidence level. Above $4 \times 10^{19}$ eV
less than $50 \%$ can be photons at the same confidence level. If we
assume the spectrum of \cite{Wolfendale} instead the bound for photon
increases (decreases) to $25\%$ ($70 \%$) at energies above
$10^{19}~$eV ($4 \times 10^{19}~$eV).

 The results for the photon bound depend on the hadronic model we choose but
 in a way that is conservative.  If we were to chose a model that produces
 fewer muons that QGSJET we would predict a composition heavier than
 protons. If we chose a model that produces more muons, we would require a
 lighter composition and more photon flux would be allowed.  From the KASCADE
 project \cite{kaska} it is evident that all models tested except for sibyll 
 produce muon rates
 above that found in the data. 
 So models that produce more muons are disfavoured. 

 Our photon bound is also conservative because we have not taken into account
 the interactions of the high energy photons in the magnetic field of the
 Earth \cite{magnetic}. This has the effect of converting a single energetic
 photon into a few lower energy photons. 
 As the total number of muons in a
 shower initiated by a single photon scales approximately with $E^{1.2}$, the 
 number of
 muons in a shower initiated by a single photon exceeds the total number of
 muons if the photon energy is split into multiple photon showers of lower energy.

The implementation of photohadronic
interactions in the AIRES code~\cite{AIRES} and CORSIKA code~\cite{CORSIKA}
(using the parameterization of \cite{Stanev}) give predictions of the total
number of muons that are equal to within 10\% at $10^{19}$~eV. 
Unless the 
photoproduction cross section has a dramatic increase at high energies, 
the photon bound is robust because the photoproduction cross section 
is small relative to hadronic interactions. 

\section{Discussion}

Conventional acceleration mechanisms, so
called ''bottom up'' scenarios, predict an extragalactic origin with
mainly proton composition as although nuclei of higher charge are more
easily accelerated they are fragile to photonuclear processes in the
strong photon fields to be expected in likely source regions
\cite{hillas_diagram}. ''Top down'' models explain the highest energy
cosmic rays arising from the decay of some sufficiently massive
``X-particles''.  These models predict particles such as nucleons,
photons and even possibly neutrinos as the high energy cosmic rays,
but not nuclei.  In some models \cite{bkv,birkel,rubin} these
X-particles are postulated as long-living metastable super-heavy relic
particles (MSRP) clustering in our galactic halo.  For these MSRP
models a photon dominated primary composition at $10^{19}$ eV is
expected. Other top down models \cite{TD} associate X-particles with
processes involving systems of cosmic topological defects which are
uniformly distributed in the universe, and predict a photon dominated
composition only above $\sim 10^{20}$~eV. 

\begin{figure}
\ybox{0.3}{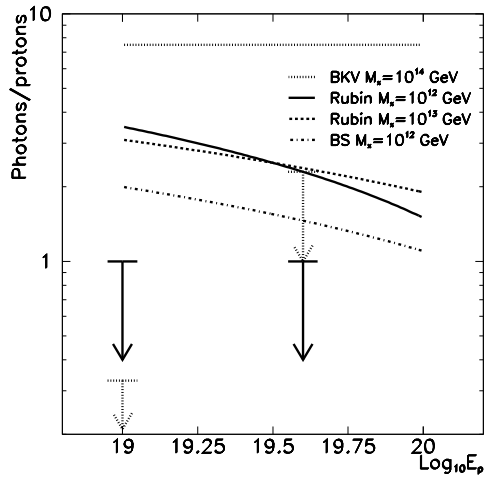}   
\caption{Photon to proton abundance ratio as a function of the energy for 
three different models for the origin of high energy cosmic rays by Berezinsky 
et al. (BKV) \cite{bkv}, Birkel et al. (BS) 
\cite{birkel}, and Rubin \cite{rubin}), and the $95 \% $ CL bounds presented in 
this work.}
\label{abund2}
\end{figure}

On general grounds dominance of 
photons over protons is expected for these models due to the QCD 
fragmentation functions of quarks and gluons from X-particle decays 
into mesons and baryons. 
The ratio of photons to protons for MSRP models is typically 10 
\cite{bkv} at 10$^{19}$ eV from QCD fragmentation. 
However some models predict a ratio closer to 2 \cite{birkel}. 
The difference depends on distance to the sources because the photons 
attenuate in shorter distances than the protons in the cosmic microwave 
background and thus can become suppressed relative to the protons if 
the sources are distant.  
Clearly, our bound on the photon flux puts severe constraints on some 
''top down'' models. This is illustrated in Fig.~\ref{abund2} where 
this ratio is plotted for three such models and compared to our bound. 

Observations above $10^{19}$~eV are otherwise consistent with both 
top down and bottom up interpretations \cite{FlysEye,agasa}. 
There is however some partial evidence against 
the photon hypothesis. Shower development of the highest energy event
\cite{FlysEye}, is inconsistent with a photon initiated shower
\cite{vazquez} while AGASA measurements of the muon lateral
distribution of the highest energy events are compatible with a proton
origin~\cite{agasacorsika}. Our result has been recently confirmed 
by comparing muon and electron densities in vertical air showers 
detected by the AGASA array \cite{Teshima01}. 

Here we have described a new method to analyse inclined showers. 
The method opens up a new way to measure cosmic ray showers. 
These showers are complementary to vertical showers because they 
are mostly due to muons that are produced far away from the detection 
point. The method can be applied to array detectors that use water 
\v Cerenkov tanks such as the Auger observatories now in construction. 

The power of analysing inclined showers is illustrated with the 
analysis of the Haverah Park data. This analysis has allowed us to set 
the first limit to the photon content of the highest energy cosmic 
rays. 
We conclude that observations of inclined showers provide a powerful tool 
to discriminate between photon and proton dominated compositions. 

{\bf Acknowledgements:} We thank Gonzalo Parente for suggestions after reading
the manuscript. This work was partly supported by a joint grant from the
British Council and the Spanish Ministry of Education (HB1997-0175), by Xunta
de Galicia (PGIDT00PXI20615PR), by CICYT (AEN99-0589-C02-02) and by
PPARC(GR/L40892).

\end{document}